\documentclass[11pt,a4paper]{article}
\usepackage{jheppub}
\usepackage[usenames,dvipsnames]{pstricks}

\newcommand{\lbl}[1]{\label{eq:#1}}
\newcommand{ \rf}[1]{(\ref{eq:#1})}

\newcommand{\be}{\begin{equation}}
\newcommand{\ee}{\end{equation}}
\newcommand{\bea}{\begin{eqnarray}}
\newcommand{\eea}{\end{eqnarray}}
\newcommand{\setl}{\setlength\arraycolsep{2pt}} 

\newcommand{\noi}{\noindent}
\newcommand{\nn}{\nonumber}
\newcommand{\ra}{\rightarrow}
\newcommand{\Ra}{\Rightarrow}

\newcommand{\cA}{{\cal A}}
\newcommand{\cB}{{\cal B}}

\newcommand{\cK}{{\cal K}}
\newcommand{\cL}{{\cal L}}
\newcommand{\cM}{{\cal M}}

\newcommand{\cO}{{\cal O}}

\newcommand{\Imm}{\mbox{\rm Im}}
\newcommand{\Ree}{\mbox{\rm Re}}

\newcommand{\MeV}{\mbox{\rm MeV}}
\newcommand{\GeV}{\mbox{\rm GeV}}

\newcommand{\annd}{\mbox{\rm and}}
\newcommand{\foor}{\mbox{\rm for}}

\newcommand{\omegah}{\hat{\omega}}

\newcommand{\Meijer}[5]{{\rm G}^{#1}_{#2} \left(#3\left|\begin{matrix} {#4} \\ {#5} \end{matrix}\right.\right) }

\title{On the Time Momentum Representation \\ of  Hadronic Vacuum Polarization and $\boldsymbol{g_{\mu}-2}$} 

\author[1]{David Greynat}
\author[2]{Eduardo de Rafael}

\affiliation[1]{No affiliation at present}
\affiliation[2]{Aix-Marseille Univ, Universit\'{e} de Toulon, CNRS, CPT, Marseille, France}

\emailAdd{david.greynat@gmail.com}
\emailAdd{EdeR@cpt.univ-mrs.fr}			

\abstract{We propose a new set of model independent approximants adapted to the time momentum representation (TMR) of hadronic vacuum polarization (HVP) and its contribution to $g_{\mu}-2$. They provide a way to extrapolate  lattice QCD (LQCD) results obtained in an optimal time-region, to the full range required for an evaluation of the HVP contribution to $g_{\mu}-2$.  They offer as well a new way to confront  LQCD results  in restricted TMR regions, with the full contribution obtained from  data driven determinations.}

\begin{document}

\makeatletter
\def\@fpheader{\relax}
\makeatother
\maketitle

\section{Introduction}

\noi
The measurements of the anomalous magnetic moment of the muon $a_{\mu}$,  made  at BNL ~\cite{E821} and more recently at Fermilab~\cite{FNAL,FL21}, give the results:
\be
a_{\mu}^{{\rm BNL}}=116~592~089(63)\times 10^{-11}\quad\annd\quad a_{\mu}^{{\rm FNAL}}=116~592~040(54)\times 10^{-11}\,.
\ee
They agree with each other at the level of  0.6 standard deviations ($0.6\sigma$) and their combined number
\be\lbl{eq:BNLFL}
a_{\mu}(2021)=116~592~040(41)\times 10^{-11}\,,
\ee
has the remarkable accuracy of 0.35 parts per million.

 The theoretical evaluation of the same observable in the Standard Model   has been made to a comparable precision. 
The result
\be
a_{\mu}({\rm Th.WP})=116~591~810(43)\times 10^{-11}
\ee
 reported in the 2020 White Paper (WP) of ref.~\cite{PhRe20} has been the {\it consensus theory number} for a while. When compared to the experimental number in Eq.~\rf{eq:BNLFL} it turns out to be  4.2$\sigma$ below, a significant difference, which has triggered many speculations  on what kind of  new physics  could explain this difference.

 The 2020 WP number, however, does not take into account the  
lattice QCD (LQCD) result of the BMW collaboration~\cite{BMWmu} 
\be\lbl{eq:LQCDL}
a_{\mu}({\rm HVP})_{\rm BMW}=7~075(55)\times 10^{-11}\,,
\ee
which differs from the evaluations using  data-driven dispersion relations~\cite{Davier,Teubner}:
\be\lbl{eq:disperevs}
a_{\mu}({\rm HVP})_{\rm lowest~order}^{\rm ref.[6]} = 6~940(40)\times 10^{-11}\quad\annd\quad a_{\mu}({\rm HVP})_{\rm lowest~order}^{\rm ref.[7]}=6~928(24)\times 10^{-11}\,,
\ee
incorporated in the
 {\it consensus theory number} of the 2020 WP. The BMW-lattice QCD result reduces the total discrepancy with the experimental result in Eq.~\rf{eq:BNLFL} from 4.2$\sigma$ to 1.6$\sigma$. Still a discrepancy, but not significant  to argue evidence for new physics.
Recently, the BMW result has also been confirmed, at least partially,  by other LQCD collaborations~\cite{wittig,jansen,khadra}. If the disagreement between LQCD and the experimental dispersive evaluations of the HVP persists, one will have to find the explanation for that. The two methods involve integrals of different quantities which makes the comparison difficult but not impossible. A lot of activity on that is underway, mostly concentrated on evaluations of the so called {\it window observables} proposed in ref.~\cite{window}~(see e.g. refs.~\cite{col22}, \cite{wittig}, \cite{RBC} and references therein).

In the meantime, the Fermilab Muon g-2 experiment expects to  reduce the error of their 2021 result by a factor of four, as more statistics accumulate. There is also a new experiment at the Japan Proton Accelerator Research Complex in Tokai, the J-PARC experiment E34~\cite{JPARC}, which will employ a new different technique to measure the muon anomaly. 
Another expected experiment is the  
 MUonE proposal  at the CERN SPS~\cite{CPTV,ABetal,MUonEP}. It   
 consists in extracting the value of the HVP self-energy  function in the Euclidean from its contribution to the differential cross-section of elastic muon-electron scattering, with muons at ${\rm E}_{\mu}=160~\GeV$ colliding on atomic electrons of a fixed low Z target~\cite{Betal}. The muon anomaly can then be obtained from a weighted integral of the measured HVP self-energy  function.

 The purpose of this paper is to introduce a new type of model independent approximants adapted to the time momentum representation (TMR) of hadronic vacuum polarization   used in  LQCD evaluations of $a_{\mu}({\rm HVP})$ at present. The method is  based on 
the {\it reconstruction approximants} which follow from the \emph{transfer theorem} of Flajolet and Odlyzko~\cite{FOth,FS09}, and has  previously been  applied to the MUonE-proposal~\cite{GdeR22} as well as to other observables (see e.g. refs.~\cite{Perisetal}).    We show how  to adapt this method  to extrapolate the LQCD results obtained in a restricted TMR-interval to the full integration domain required to  evaluate  $a_{\mu}({\rm HVP})$.

The paper is organized as follows. Section \ref{Sect:II} reviews  the properties of HVP and its TMR which will   be needed. Section \ref{Sect:III} is dedicated to the asymptotic behaviours of the TMR function $G(x_0)$ in QCD,  both at short distances  and at long  distances. As far as we know, some aspects of this section are new, in particular the construction of a skeleton $G^* (x_0)$ function in terms of Bessel functions which provides  a first approximant to the TMR $G(x_0)$ function in its full $x_0$ range. Section \ref{Sect:IV} discusses the formulation of the  reconstruction approximants that follow from  the transfer theorem of Flajolet and Odlyzko~\cite{FOth,FS09}. The content of this theorem is explained in Subsection \ref{Sect:IV1}, and its application to construct what we call FO-approximants (for short) is discussed in detail in Subsection \ref{Sect:IV2}. Section \ref{Sect:V} is dedicated to show how to implement the FO-approximants in practice, and we illustrate this  with the example of a phenomenological model which simulates the physical hadronic spectral function. The conclusion and outlook are finally given in Section \ref{Sect:VI}. We have relegated to an Appendix the mathematical details of the FO-theorem  needed in  our application.

\section{Properties of HVP and its TMR }\label{Sect:II}

The function which governs HVP is the Fourier transform of the vacuum expectation value of the time-ordered product of two electromagnetic hadronic  currents of the Standard Model $J_{\mu}^{\rm had}(x)$ at separate space-time $x$-points:

\be\lbl{eq:ft}
\Pi_{\mu\nu}^{\rm had}(q)=i\int_{-\infty}^{+\infty}\,  d^4 x\  e^{iq\cdot x}
\langle 0\vert T\left(J_{\mu}^{\rm had}(x)J_{\nu}^{\rm had}
(0)\right)\vert 0\rangle=(q_{\mu}q_{\nu}-q^2 g_{\mu\nu})\Pi_{\rm had}(q^2)\,.
\ee
The  hadronic photon self-energy function $\Pi_{\rm had}(q^2)$  is a complex function of its  $q^2$ variable. It is an analytic function  in the full complex plane, but for a cut in the real axis which  goes  from the physical threshold $t_0$ to infinity~\footnote{In the presence of higher order electromagnetic corrections the threshold is at the mass of the $\pi^0$ because of the $\pi^0 \gamma$ contribution to the spectral function. In this  paper the threshold will be fixed at $t_0 =4m_{\pi^\pm}^2$, but can be adjusted to $m_{\pi^0}^2$ if necessary.}. As such, the on-shell renormalized HVP-function,  i.e. $\Pi_{\rm had}(q^2)$ subtracted at its value at $q^2 =0$,  obeys the dispersion relation: 
\be\lbl{eq:Pi}
\Pi^{\rm HVP}(q^2)\equiv \Pi_{\rm had}(q^2)-\Pi_{\rm had}(0) =  \int_{t_0}^{\infty}\frac{dt}{t}\,
\frac{q^2}{t-q^2 -i\epsilon}\frac{1}{\pi}\Imm\Pi_{\rm had}(t)\,,\quad t_0 \equiv 4m_{\pi^\pm}^2\,,
\ee
and the optical theorem relates the hadronic spectral function $\frac{1}{\pi}\Imm\Pi_{\rm had}(t)$ to the  one-photon annihilation cross-section into hadrons:
\be
\sigma(t)_{e^+ e^- \ra {\rm had}} \underset{{m_e\ra 0}}{\thicksim}\frac{4\pi^2 \alpha}{t}\frac{1}{\pi}\Imm\Pi_{\rm had}(t)\,.
\ee
The evaluation of  the HVP contribution to the anomalous magnetic moment of the muon $a_{\mu}^{\rm HVP}$ can then be made using the integral representation~\cite{BM61,BdeR,GdeR}:
\be\lbl{eq:BMSR}
a_{\mu}^{\rm HVP}=\frac{\alpha}{\pi}\int_{t_0}^\infty \frac{dt}{t}\int_0^1 dx\  \frac{x^2 (1-x)}{x^2 +\frac{t}{m_{\mu}^2}(1-x)}\frac{1}{\pi}\ \Imm\Pi_{\rm had}(t)\,.
\ee
This so called {\it dispersive method}, is the way that experimental data-driven determinations of $a_{\mu}^{\rm HVP}$ have been made;  the earliest  in ref.~\cite{GdeR} using the Gounaris-Sakurai parametrization of the pion form factor~\cite{GS68}, the latest in refs.~\cite{Davier, Teubner} using a wealth of experimental results.

A crucial observation made by the authors of ref.~\cite{BM11} is that, in Euclidean space-time 
and in the special kinematic configuration where $\vec{q}=0$, the 
$\Pi_{\mu\nu}^{\rm had}(q)$ function in Eq.~\rf{eq:ft} becomes 
\be\lbl{eq:under}
\Pi_{ij}^{\rm had}(q_0,\vec{0})=\int_{-\infty} ^{+\infty} dx_0 e^{-iq_0 x_0}\underbrace{ \int_{-\infty} ^{+\infty} d^3 \vec{x}\  \delta_{ij}\  \langle 0\vert T\left(J_{i}(x_0, \vec{x})J_{j}
(0)\right)\vert 0\rangle}_{G(x_0)} \,,
\ee
and the underlined time-dependent function $G(x_0)$, for $x_0$ in  an optimal region, is accessible to accurate evaluations in LQCD. The expression  of $a_{\mu}^{\rm HVP}$ in terms of $G(x_0)$, the so called TMR~\cite{BM11}, is then  given by the integral:
\begin{multline}
\lbl{eq:ETRNjd}
a_{\mu}^{\rm HVP} = \frac{\alpha}{\pi} m_{\mu}^2\int_0^\infty  d{x}_0 \ \frac{x_0^4}{2}
\;\Meijer{2,3}{3,5}{(m_{\mu}{x}_0)^2}{-1, -\frac{1}{2}, 0 \,; \,\relbar\!\relbar}{ 0, 1, -3, -\frac{3}{2}, -2} \\
\times \underbrace{\int_{\sqrt{t_0}}^\infty d\omega\  \omega^2\   e^{-\omega\vert{x}_0\vert
}\frac{1}{\pi}\Imm\Pi_{\rm had}(\omega^2)}_{G({x}_0)}\,, 
\end{multline}
where $\omega^2=t$ (the Minkowski  $t$-variable of the spectral function) and 
\begin{multline}
\lbl{eq:MNRGF0}
\Meijer{2,3}{3,5}{(m_{\mu}{x}_0)^2}{-1, -\frac{1}{2}, 0 \,; \,\relbar\!\relbar}{ 0, 1, -3, -\frac{3}{2}, -2}\\
= \frac{1}{2\pi i} \int\limits_{c_s -i\infty}^{c_s +i\infty}ds\  (m_{\mu}{x}_0)^{-2s} 
\frac{\Gamma(s)\Gamma(1+s)\Gamma(1-s)\Gamma(\frac{3}{2}-s)\Gamma(2-s)}{\Gamma(4-s)\Gamma(3-s)\Gamma(\frac{5}{2}-s)}
\,,
\end{multline}
is a  Meijer's G-function.

The TMR-function $G(x_0)$  in Eq.~\rf{eq:ETRNjd} is the second derivative (with respect to the time variable  $x_0$) of the Laplace transform (with respect to the energy variable $\omega$) of the hadronic spectral function. From  the usual definition of the Laplace transform:
\be\lbl{eq:LapTG}
\cL(x_0)=\int_{\sqrt{t_0}}^\infty d\omega\  e^{-w x_0}\ \frac{1}{\pi}\Imm\Pi_{\rm had}(\omega^2)\,,
\ee 
there follows that
\be\lbl{eq:LapTG2}
G(x_0)=\left(-\frac{\partial}{\partial x_0}\right)^2 \cL(x_0)= 
\int_{\sqrt{t_0}}^\infty d\omega\   e^{-\omega{x}_0
}\ \omega^2 \frac{1}{\pi}\Imm\Pi_{\rm had}(\omega^2)\,.
\ee
Because of the positivity of the hadronic spectral function, both functions $\cL(x_0)$ and  $G(x_0)$ as well as the successive derivatives $\left(-\frac{\partial}{\partial x_0}\right)^p G(x_0)\,, p=1,2,3,\dots$, are all monotonously decreasing functions of $x_0$ for $0\le x_0 \le \infty$; a well known property as well of the Mellin transform of the hadronic spectral function~\cite{ChGdeR18}
\be\lbl{eq:MBSHF}
\cM(s)=\int_{t_0}^\infty \frac{dt}{t}\left(\frac{t}{t_0}\right)^{s-1}\frac{1}{\pi}\Imm\Pi_{\rm had}(t)\,,
\ee
as a function of $\Ree(s)<1$.

An alternative way to evaluate $a_{\mu}^{\rm HVP}$ to the one in Eq.~\rf{eq:ETRNjd},  is to use
 the integral representation
\be
\Pi^{\rm HVP}(-Q^2) = 2\int_0^\infty dx_0\  [1-\cos(\sqrt{Q^2} x_0)]\  \cL (x_0)\,,
\ee 
 equivalent to~\cite{BM11}
\be
\Pi^{\rm HVP}(-Q^2) = 2 \int_0^\infty dx_0 \left[\frac{1-\cos(\sqrt{Q^2} x_0)}{Q^2}-\frac{x_0^2}{2}\right]\ G(x_0)\,,
\ee 
and then the Euclidean representation of the anomaly  proposed in refs.~\cite{LPdeR,EdeR94,Blum}:
\be\lbl{eq:LdeR}
a_{\mu}^{\rm HVP} 
 =  -\frac{\alpha}{\pi}\int_0^1 dx\ (1-x)\ \Pi^{\rm HVP}\left( -\frac{x^2}{1-x}m_{\mu}^2\right)\,,\quad Q^2 \equiv \frac{x^2}{1-x}m_{\mu}^2\,.
\ee

 \section{Asymptotic Behaviours}\label{Sect:III}

The TMR-function $G(x_0)$ has a Mellin-Barnes representation which can be obtained by inserting the identity
\be
e^{-\omega x_0}=\frac{1}{2\pi i}\int\limits_{c_s -i\infty}^{c_s +i\infty}ds\ {(\omega x_0)}^{-s}\ \Gamma(s)\,, \quad c_s \equiv\Ree(s)> 0\,,
\ee
in the integrand of Eq.~\rf{eq:LapTG2} (recall that $t=\omega^2$) and following the steps:

{\bea\lbl{eq:MBRG}
G(x_0) & = & 
\int_{\sqrt{t_0}}^\infty d\omega\   e^{-\omega{x}_0
}\ \omega^2\  \frac{1}{\pi}\Imm\Pi_{\rm had}(\omega^2)\\ 
& = & \frac{1}{2\pi i}\int\limits_{c_s -i\infty}^{c_s +i\infty}ds\ x_0^{-s}\ \Gamma(s)  \frac{1}{2}\int_{t_0}^\infty \frac{dt}{t}\ t^{3/2 -s/2} \frac{1}{\pi}\Imm\Pi_{\rm had}(t)\quad\quad (t\equiv \omega^2 ) \nn \\ 
 & = &  \frac{t_0^{3/2}}{2}\frac{1}{2\pi i}\int\limits_{c_s -i\infty}^{c_s +i\infty}ds\ (x_0 \sqrt{t_0})^{-s}\ \ \Gamma(s)\ \cM(5/2-s/2)\,, \quad c_s \equiv\Ree(s)> 3 \nn \\
 & = & t_0^{3/2}\ \frac{1}{X^5}\ \frac{1}{2\pi i}\int\limits_{c_s -i\infty}^{c_s +i\infty}ds \left(\frac{1}{X^2}\right)^{-s} \Gamma(5-2s)\ \cM(s)\,, \quad c_s \equiv\Ree(s)<1\,, 
\lbl{eq:MBG} 
\eea}
where going from the second line to the third  we have used the definition of the Mellin transform of the spectral function in Eq.~\rf{eq:MBSHF}, and  from the third line to the fourth  we have introduced the dimensionless variable
\be\lbl{eq:xscaled}
X\doteq x_0\sqrt{t_0}\,,
\ee
and redefined the integration $s$-variable. From here onwards we shall often work with 
  the  dimensionless TMR-$G(X)$ function: 
\be\lbl{eq:Gdim}
G(X)\equiv \int_1^\infty d\omegah\ e^{-\omegah X}\omegah^2 \frac{1}{\pi}\Imm\Pi_{\rm had}(\omegah^2 t_0)= \frac{1}{t_0^{3/2}}G(x_0)\,,
\ee
where
\be
\omegah\equiv \frac{\omega}{\sqrt{t_0}}\,.
\ee
The TMR of the muon anomaly with this redefinition in terms of the dimensionless variable $X$ is then:
\be\lbl{eq:TMRdiml}
a_{\mu}^{\rm HVP} = \frac{\alpha}{\pi} \frac{m_{\mu}^2}{t_0}\int_0^\infty  dX \ {\cK}(X)\    X^3\  G(X)\,, 
\ee
with
\be\lbl{eq:kernel}
{\cK}(X)= \frac{X}{2}\;\Meijer{2,3}{3,5}{\frac{m_{\mu}^2}{t_0}\ X^2}{-1, -\frac{1}{2}, 0 \,; \,\relbar\!\relbar}{ 0, 1, -3, -\frac{3}{2}, -2}
\ee
the integration kernel. 
Figure \rf{fig:Fig.1} shows the familiar  shape of this kernel, as a function of $X$ and  as a function of $x_0$ in Fermi units for comparison.
\vspace*{0.25cm}
\begin{figure}[!ht]
\begin{center}
\hspace*{0.25cm}
\includegraphics[width=0.50\textwidth]{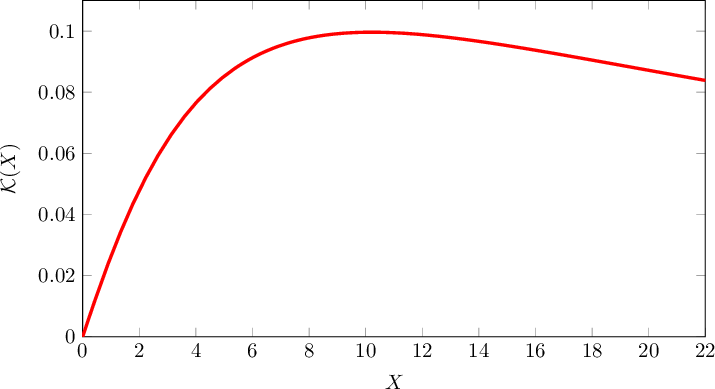}\;\includegraphics[width=0.50\textwidth]{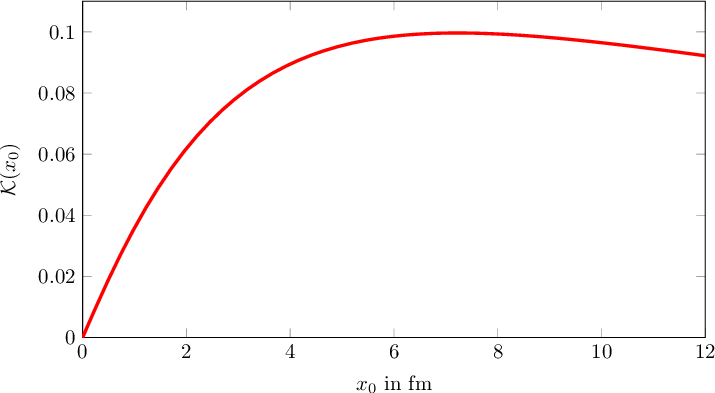}
\caption{Plots of the kernel $\mathcal{K}(X)$ in Eq.~\rf{eq:kernel} versus $X$ and versus $x_0$ in Fermi units.}
\lbl{fig:Fig.1}
\end{center}
\end{figure}

The Mellin-Barnes representation of $G(X)$ that follows from Eq.~\rf{eq:MBG} is:
\be\lbl{eq:MBGX}
G(X)=\frac{1}{X^5}\ \frac{1}{2\pi i}\int\limits_{c_s -i\infty}^{c_s +i\infty}ds \left(\frac{1}{X^2}\right)^{-s} \Gamma(5-2s)\ \cM(s)\,, \quad c_s \equiv\Ree(s)<1 \,,
\ee
where the QCD  dynamics is fully encoded in the Mellin transform of the hadronic spectral function $\cM(s)$ defined in Eq.~\rf{eq:MBSHF}. 
The {\it singular expansion}~\cite{FGD} of the $s$-integrand in this representation produces the following series expansion  for $0\le X \le 1$  ( {\it i.e.  short distances}):
\be\lbl{eq:xex}
G(X) \underset{X\ra 0} {\sim}\ \frac{\alpha}{\pi}\ \left\{\frac{a_{-3}}{X^3}+ \frac{a_{-1}}{X} +  \sum_{n=1}^\infty \left[ a_{n} +b_{n}\log X \right] X^n \right\}  \,.
\ee
The  coefficient $a_{-3}$  of the leading term	is 
 fixed  by the residue of the QCD Mellin transform  at $s=1$ which, to leading order in pQCD, is
\be\lbl{eq:pQCD}
a_{-3}=\frac{N_c}{3}\sum_{\rm quarks}e_q^2 \times 2\,.
\ee
 The next coefficient $a_{-1}$ is  governed by the quark mass terms of  $\cO\left(1/Q^2\right)$ in the expansion of  $\Pi_{\rm had}(-Q^2)$ at large $Q^2$. The contribution from the light quark masses to  $a_{-1}$ vanishes in the chiral limit. The form of the rest of the asymptotic series  in Eq.~\rf{eq:xex} assumes  that the singularities at $s=1,2,3,\cdots$ of $\cM(s)$ are simple poles,    otherwise higher power  $\log X$-terms must also be included. Let us recall (see e.g. ref~\cite{GdeR22}) that  the singularities of $\cM(s)$ at $s=1,2,3,\dots$  govern the asymptotic expansion of the hadronic self-energy at large $Q^2$. The coefficients $a_n$ and $b_n$, however, will become  free parameters in our approach; only the value of $a_{-3}$ in Eq.~\rf{eq:pQCD} will be used as an input.

The Mellin-Barnes representation in Eq.~\rf{eq:MBGX} does not give, however, direct  information about the behaviour of $G(X)$ at large-$X$ ({\it i.e. long-distances}). This is because the {\it fundamental strip}~\cite{FGD} where the integral in Eq.~\rf{eq:MBGX} converges goes all the way from $\Ree(s)<1$ to $\Ree(s)=-\infty$. One can nevertheless show, using inverse Laplace-transform  properties~\cite{olver74}, that the large-$X$ behaviour of $G(X)$  is related to the $\hat{\omega}\ra 1$ threshold behaviour of the hadronic spectral function, i.e. to the power series:  
\be\lbl{eq:thexp}
\hat{\omega}^2\ \frac{1}{\pi}\Imm\Pi_{\rm had}(\hat{\omega}^2 t_0) \underset{\hat{\omega}\ra 1} {\sim}
\frac{\alpha}{\pi}\ (\hat{\omega}-1)^{3/2} \sum_{n=0}^\infty \chi_n (\hat{\omega}-1)^n\,,
\ee
where, to lowest order in chiral perturbation theory ($\chi$PT)
\be\lbl{eq:chizero}
\chi_{0}\Ra \frac{1}{3\sqrt{2}}\,,\quad \chi_{1} \Ra \frac{-1}{12\sqrt{2}}\,,\quad \chi_2 \Ra \frac{11}{96\sqrt{2}}\,, \quad \cdots\,.
\ee
  Higher order $\chi$PT corrects these values by a series in threshold $t_0$-powers:
\be\lbl{eq:chi0c}
\chi_0 \Ra \frac{1}{3\sqrt{2}}\left(1+\frac{1}{3}\langle{\rm r}^2\rangle_{\pi^{\pm}}\ t_0 +\cdots \right)\,,\quad 
\chi_1 \Ra \frac{-1}{12\sqrt{2}}\left(1-\frac{7}{3}\langle{\rm r}^2\rangle_{\pi^{\pm}}\ t_0 + \cdots \right)\quad\cdots\,,
\ee
where e.g., at the one loop level in $\chi$PT~\cite{GL85b}
\be
\langle{\rm r}^2\rangle_{\pi^{\pm}}=\frac{12 {\rm L}_{9}(\mu)}{f_{\pi}^2}- \frac{1}{32\pi^2 f_{\pi}^2}\left[2\log\left(\frac{m_{\pi}^2}{\mu^2}\right)+\log\left(\frac{m_{K}^2}{\mu^2} \right) +3 \right]\,,
\ee
and the low-energy constant ${\rm L}_{9}(\mu)$ can be obtained, either from experiment~\cite{MSR}:
\be
\langle{\rm r}^2\rangle_{\pi^{\pm}}=(0.439\pm 0.008)\ {\rm fm}^2\quad\Ra\quad {\rm L}_{9}(M_{\rho})=(6.9\pm 0.7)\times 10^{-3}\,,
\ee
or  from LQCD determinations which are in good agreement (see e.g. ref.~\cite{FLAG})  with the experimental value.

Numerically
\be\lbl{eq:chi0}
\chi_0 =\frac{1}{3\sqrt{2}}\left(1+\frac{1}{3}\langle{\rm r}^2\rangle_{\pi^{\pm}}\ t_0\right)= 0.284\pm 0.001\,,
\ee
and
\be 
\chi_1 = \frac{-1}{12\sqrt{2}}\left(1-\frac{7}{3}\langle{\rm r}^2\rangle_{\pi^{\pm}}\ t_0\right) = 0.026\pm 0.002\,.
\ee

 Inserting the threshold expansion in the integrand of the $G(X)$  function in Eq.~\rf{eq:Gdim} leads to  the {\it long-distance} behaviour of $G(X)$ in terms of a series of simple Laplace transforms:

{\setl
\bea
G(X) & \underset{X\ra \infty} {\sim} & \ \frac{\alpha}{\pi}\  \sum_{n=0}^\infty \chi_{n}\ e^{-X}\ \int_{1}^\infty d\hat{\omega}\  e^{-(\hat{\omega}-1)X}   (\hat{\omega}-1)^{n+3/2}\lbl{eq:LDS} \\
& = & \ \frac{\alpha}{\pi}\ \frac{e^{-X}}{X^{\frac{5}{2}}}\ \sum_{n=0}^\infty \chi_{n}\ \Gamma\left(\frac{5}{2}+n\right) \frac{1}{X^n}\,. \lbl{eq:ldGX}
\eea}

\noi
 This series, however, is a divergent series (though Borel summable in all the models we have examined), which implies that both the small-$X$ expansion as well as the large-$X$ expansion of $G(X)$ will participate in the  application of the transfer theorem of Flajolet and Odlyzko~\cite{FOth,FS09} that we shall later discuss.

 \subsection{Asymptotic Expansions of $G(X)$ and   Bessel Functions} \label{Sect:III1}

An interesting observation about the {\it long distance} behaviour of $G(X)$ in Eq.~\rf{eq:LDS}  is that each term of the series in the r.h.s. can be expressed as a sum of  modified Bessel functions of the second kind ${\rm K}_{n}(X)$ (BesselK[n,X] in Wolfram's Mathematica notation). This follows from the integral representation~\footnote{ See e.g. ref.~\cite{NIST:DLMF}}:
\be\lbl{eq:BIR}
{\rm K}_{n}(X)=\frac{2^{-n}\sqrt{\pi}}{\Gamma(n +1/2)}X^n \int_{1}^\infty d\hat{\omega}\ e^{-X \hat{\omega}}
(\hat{\omega}^2 -1)^{n-1/2}\,,
\ee
and the fact that the series expansion in Eq.~\rf{eq:thexp} can be rearranged as follows:
\be\lbl{eq:rearr}
\hat{\omega}^2\ \frac{1}{\pi}\Imm\Pi_{\rm had}(\hat{\omega}^2 t_0) \underset{\hat{\omega}\ra 1} {\sim}\ 
\frac{\alpha}{\pi}\ (\hat{\omega}^2-1)^{3/2} \sum_{n=0}^\infty \hat{\chi}_n (\hat{\omega}^2-1)^n\,,
\ee
with the coefficients $\hat{\chi}_n$ recursively related to the $\chi_n$-coefficients in Eq.~\rf{eq:ldGX}: 
\begin{align}
&\chi_0=2\sqrt{2}\hat{\chi}_0\,,\quad 
\chi_1 =\frac{11}{\sqrt{2}}\hat{\chi}_0 +4\sqrt{2} \hat{\chi}_{1}\,,\nonumber\\
&\chi_2 =\left(\frac{3}{8\sqrt{2}}+5\sqrt{2}\right)\hat{\chi}_1 +13\sqrt{2}\hat{\chi}_2 +8\sqrt{2}\hat{\chi}_3\,,\quad \cdots
\end{align}
The equivalent {\it long-distance} asymptotic behaviour of $G(X)$ in  terms of the $\hat{\chi}_n$-series in Eq.~\rf{eq:rearr} is then
\be
G(X)  \underset{X\ra \infty} {\sim} \ \frac{\alpha}{\pi}\  \sum_{n=0}^\infty \hat{\chi}_{n}\ e^{-X} \frac{\partial^2}{\partial X^2}\left( \int_{1}^\infty d\hat{\omega}\ e^{-X \hat{\omega}}
(\hat{\omega}^2 -1)^{n+3/2}\right)
\ee
which, using the integral representation of the Bessel function in Eq.~\rf{eq:BIR}, becomes
\be\lbl{eq:partser}
G(X)  \underset{X\ra \infty} {\sim} \ \frac{\alpha}{\pi}\ \sum_{n=0}^\infty \hat{\chi}_{n}\ e^{-X}\ \frac{  2^{n+2}\Gamma(n+5/2)}{\sqrt{\pi}}\ 
\frac{\partial^2}{\partial X^2}\left[\frac{1}{X^{n+2}} {\rm K}_{n+2}(X)\right]\,,
\ee
and
\begin{multline}
\frac{\partial^2}{\partial X^2}\left[\frac{1}{X^{n+2}} {\rm K}_{n+2}(X)\right]=\frac{1}{X^{5+n}}[(20 +  18n + 4n^2)X +X^3] {\rm K}_{n}(X) \\
+{  \frac{1}{X^{5+n}}}[40 + 76n +44 n^2 +8 n^3 + (7+4n) X^2]\ {\rm K}_{n+1}(X)\,.
\end{multline}
In particular, the  $n=0$ term  of the series in Eq.~\rf{eq:partser} is
\be
3\hat{\chi}_0\ \frac{e^{-X}}{X^5}\left[(20X + X^3){\rm K}_0 (X)+(40 +7X^2){\rm K}_1 (X)\right]
\ee
which, using the fact that
\be\lbl{eq:LXB}
{\rm K}_n (X) \underset{X\ra \infty} {\sim} e^{-X}\ \sqrt{\frac{\pi}{2}} \left(\frac{1}{X^{1/2}}+ \frac{4n^2 -1}{8} \frac{1}{X^{3/2}}+ \cdots\right)\,,
\ee
reproduces the leading behaviour of $G(X)$ at $X\ra\infty$ in Eq.~\rf{eq:ldGX} because:
\be
3\hat{\chi}_0\ \frac{e^{-X}}{X^5}\left[(20X + X^3){\rm K}_0 (X)+(40 +7X^2){\rm K}_1 (X)\right]\underset{X\ra \infty} {\sim} 3\hat{\chi}_0 \sqrt{\frac{\pi}{2}}\  \frac{e^{-X}}{X^{5/2}}\,.
\ee

Another observation  about Bessel functions concerns the asymptotic behaviour of $G(X)$ at small-$X$. It is the fact that  ${\rm K}_0 (X) \underset{X\ra 0}{=} \mathcal{O}(\ln X)$ and ${\rm K}_\nu (X) \underset{X\ra 0} {=} \mathcal{O}(X^{-\nu})$ for $\nu\not= 0$ and in particular 
\be
{\rm K}_3 (X) \underset{X\ra 0} {\sim} 
\frac{8}{X^3}-\frac{1}{X}+\frac{X}{8}+
\frac{X^3}{576}(12 \log X -11+12 \gamma_{\rm E}-12 \log 2) +\cdots\,,
\ee
which has the same $\cO (X^{-3})$ leading behaviour as $G(X)$ in QCD at $X\ra 0$. 

These observations about Bessel functions  suggest considering the minimal linear combination of ${\rm K}_{n}(X)$-functions, modulated by $X$-polynomials,  which    reproduces the leading asymptotic behaviours of $G(X)$ in QCD, both at short distances and at long-distances. We  call this the {\it skeleton $G^{*}(X)$ approximant}  of the physical $G(X)$ and discuss its construction in the next subsection.  

 \subsection{The Skeleton  $G^{*}(X)$ Function}\label{Sect:III2}

The function in question must be of the form:
\be\lbl{eq:Gstar}
  G^{*}(X) =\frac{\alpha}{\pi}\ \left[ a_{-3} \frac{1}{8} {\rm K}_3(X) + \left(A_{1,0} + A_{1,1} X\right){\rm K}_1(X)
+ \left(A_{0,0} + A_{0,1} X\right){\rm K}_0(X)\right]\,,
\ee
with  $a_{-3}$ the same coefficient as in Eq.~\rf{eq:pQCD},  and the coefficients $A_{i,j}$ adjusted so as to reproduce the leading threshold behaviour of $G(X)$ in Eq.~\rf{eq:ldGX}. This requires a set of constraints on the $A_{i,j}$ coefficients so that
\be\lbl{eq:GXSL}
G^* (X) \underset{X \rightarrow \infty}{\sim}\ \frac{\alpha}{\pi}\ \frac{3}{4}\sqrt{\pi}\ \chi_0\  \frac{e^{-X}}{X^{5/2}}\,.
\ee
The constraints follow from  the fact that the $ G^{*}(X)$ function in Eq.~\rf{eq:Gstar} has the asymptotic expansion:
\begin{align}
 \frac{\pi}{\alpha}\ \sqrt{ \frac{2}{\pi}} \ e^{X}G^*(X)\  \underset{X \rightarrow \infty}{=} & \left[A_{0,0}+A_{1,1}\right] \sqrt{X} \nn \\
    &+  \frac{1}{8}\left[a_{-3}+8 A_{0,0}-A_{0,1}+8 A_{1,0}+3 A_{1,1}\right] \frac{1}{\sqrt{X}}\nn \\
    &+ \frac{1}{128}\left[70 a_{-3}-16 A_{0,0}+9 A_{0,1}+48  A_{1,0}-15 A_{1,1}\right] \frac{1}{X^\frac{3}{2}} \nn \\
    &+\frac{3}{1024}\left[315 a_{-3}+ 24 A_{0,0}-5\left(5  A_{0,1}+ 8 A_{1,0}-7 A_{1,1}\right)\right] \frac{1}{X^\frac{5}{2}}\nn\\
    &+\cO\left[\frac{1}{X^{\frac{7}{2}}} \right]\,,
  \end{align}
which, in order to agree with the leading behaviour in Eq.~\rf{eq:GXSL}, forces the coefficients of the first three terms of this expansion  to vanish and the coefficient of the  fourth term to reproduce the result in Eq.~\rf{eq:GXSL}. The solution of this system of four linear equations with four unknowns is unique and the $G^{*}(X)$ function is then completely determined in terms of the two parameters $a_{-3}$ and  $\chi_0$ with the result
\bea\lbl{eq:Gstarexp}
  G^{*}(X) & = & \frac{\alpha}{\pi}\ \left\{ \frac{a_{-3}}{8}\ {\rm K}_3(X) \right. \nn \\
		& + &  
 \left[-\frac{33}{8}\ a_{-3}+2\sqrt{2}\ \chi_0  -(12\  a_{-3}-8\sqrt{2}\ \chi_0)X\right]	{\rm K}_1(X) \nn \\
& + &\left. \left[10\  a_{-3}-6\sqrt{2}\ \chi_0 + (12 \ a_{-3} -8\sqrt{2}\ \chi_0 )X \right]{\rm K}_0(X)\right\} \,.
\eea
The  shape of the function $X^3\ G^{*}(X)$ (in $\frac{\alpha}{\pi}$ units) , for $a_{-3}= 10/3$ and the central value $\chi_0=0.28$  in Eq.~{\rf{eq:chi0}, is  
 shown in Fig.~\rf{fig:Fig.2}. As expected, it is a monotonic decreasing function. 
A quality test of the {\it skeleton interpolating approximant} $G^{*}(X)$ is  its contribution to the muon anomaly.  The result ($\mathcal{K}(X)$ is the same  kernel as in Eq.~\rf{eq:kernel})
\be\lbl{eq:TMR_Gstarl}
a_{\mu}^{\rm HVP}[G^{*}] = \frac{\alpha}{\pi} \frac{m_{\mu}^2}{t_0}\int_0^\infty  dX \ {\cK}(X)\    X^3 G^{*}(X) =7~533 \times 10^{-11} 
\ee
reproduces the central value of, e.g.  the LQCD determination in Eq.~\rf{eq:LQCDL}, at the 6\% level,  not bad for a first approximation  to the physical $G(X)$. 

\begin{figure}[!ht]
\begin{center}
\includegraphics[width=0.70\textwidth]{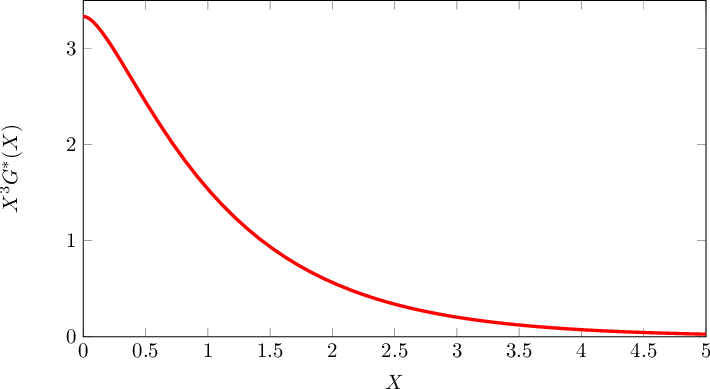} 
\caption{Plot in $\frac{\alpha}{\pi}$ units of the {\it skeleton function}  $X^3 \ G^{*}(X)$ in Eq.~\rf{eq:Gstarexp} versus $X$.}
\lbl{fig:Fig.2}
\end{center}
\end{figure}

We have also evaluated analytically the associated spectral function to $G^{*}(X)$, i.e. the {\it skeleton spectral function} $\frac{1}{\pi}\Imm \Pi^{*}(t)$ such that
\be
G^{*}(X)= \int_1^\infty d\omegah\ e^{-\omegah X}\omegah^2 \frac{1}{\pi}\Imm\Pi^{*}(\omegah^2 t_0)\,.
\ee
The derivation follows from the   analytic properties of the Bessel functions which define $G^{*}(X)$ with the result 
\be\lbl{eq:Gstarspc}
   \frac{1}{\pi} \Imm \Pi^* (t=\hat{\omega}^2  t_0) = \frac{\alpha}{\pi}\ \frac{(\hat{\omega}-1)^2\ \left[a_{-3}\left(\hat{\omega}-1\right)\ \left(\hat{\omega}+4\right)+4\sqrt{2}\ \chi_0 \right]}{2\ \hat{\omega}^2 \ (1+\hat{\omega})\ \sqrt{\hat{\omega}^2 -1} }\,.
\ee
The shape of this spectral function plotted in Fig.\rf{fig:Fig.3} (in $\frac{\alpha}{\pi}$ units)  shows  a smooth interpolation of the two asymptotic leading behaviours of the HVP spectral function:
\begin{align}
& \frac{1}{\pi} \Imm \Pi^*(t=\hat{\omega}^2 t_0) \underset{\hat{\omega}\  \rightarrow 1}{\sim}  \  \frac{\alpha}{\pi}\  \frac{1}{12}(\hat{\omega}-1)^{\frac{3}{2}}\left(1+\frac{1}{3}\langle{\rm r}^2\rangle_{\pi^{\pm}}\ t_0\right)\,, \nn \\
&\frac{1}{\pi} \Imm \Pi^*(t=\hat{\omega}^2 t_0) \underset{\hat{\omega} \rightarrow \infty}{\sim}\ \frac{\alpha}{\pi}\ \frac{5}{3} \,.
\end{align}

\begin{figure}[!ht]
\begin{center}
\includegraphics[width=0.70\textwidth]{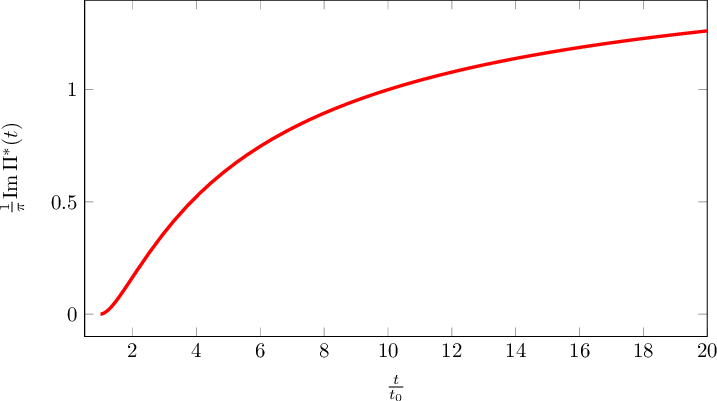} 
\caption{Plot in $\frac{\alpha}{\pi}$ units of the {\it skeleton spectral function}  $\frac{1}{\pi} \Imm \Pi^* (t)$ in Eq.~\rf{eq:Gstarspc} versus $\tfrac{t}{t_0}$.}
\lbl{fig:Fig.3}
\end{center}
\end{figure}

Our  choice of a {\it skeleton function}  is of course not unique. Any  monotonically decreasing function that interpolates  the leading asymptotic behaviours of $G(X)$ at long and short distances in QCD is  a possible choice. One may even choose as a {\it skeleton $G^{*}(X)$ function} the one provided by the data-driven determination of the HVP spectral function, as suggested in the Outlook. Our choice, however,  is good enough to implement the approximants that we discuss in the next section. These  approximants do not depend  on the choice of the {\it skeleton function}, only the speed of their convergence depends.

 \section{Flajolet-Odlyzko  Approximants}\label{Sect:IV}
We next discuss how to improve on the {\it skeleton approximant} $G^{*}(X)$ that we have chosen. The  function 
\be\lbl{eq:FOeq}
G_{\rm FO}(X)\equiv \frac{G(X)}{G^{*}(X)}-1
\ee
defines  the deviation of the hadronic  $G(X)$ function that we want to reconstruct from the chosen   skeleton $G^{*}(X)$ function. The reason why we  introduce 
 this  $G_{\rm FO}(X)$ function is that it no longer has an  exponential behaviour at long-distances  and, therefore, it is better   adapted to an application of  the transfer theorem of Flajolet and Odlyzko~\cite{FOth} (FO-theorem for short, hence the subscript FO in $G_{\rm FO}(X)$). Given some values of $G(X)$ in a finite $X$ region (i.e. a finite $x_0$ region),  we shall first apply the FO-theorem to reconstruct the corresponding $G_{\rm FO}(X)$ function in its full $x_0$ range and then, from this reconstruction,  the one of the  $G(X)$ function will follow from  Eq.~\rf{eq:FOeq}. 
 
The  asymptotic expansions of $G_{\rm FO}(X)$ can be deduced from the fact that we know  ${G^{*}(X)}$ explicitly, as well as the parametrizations of the  expansions at small-$X$ and large-$X$  of $G(X)$, with the results
\begin{align}\lbl{eq:FOseries}
  G_{\rm FO}(X) &\underset{X \rightarrow 0}{\sim} \sum_{\ell\geqslant0, n\geqslant2+\ell} \operatorname{s}_{n,\ell} X^n \ln^\ell X &\text{and}&& G_{\rm FO}(X)\underset{X \rightarrow \infty}{\sim} \sum_{n\geqslant1} \frac{\operatorname{l}_{n}}{X^n}\,.
  \end{align}
where e.g.
\begin{align}
  &\operatorname{s}_{2,0} = \frac{a_{-1}-2 \sqrt{2} \chi_0}{a_{-3}}+\frac{17}{4}\;,\\
  &\operatorname{s}_{3,0} =\frac{a_0+2 a_{-3} (6+5 \gamma_E -5 \ln 2)+2 \sqrt{2} \chi_0 (-4-3 \gamma_E +\ln 8)}{a_{-3}}\;,\\
  &\operatorname{s}_{3,1} = 10-\frac{6 \sqrt{2} \chi_0}{a_{-3}} \;, \lbl{eq:s31}\\
	& \cdots & \nn \\
  &\operatorname{l}_{1} =\frac{5 \left(-5 \sqrt{2} a_{-3}+6 \chi_0+8 \chi _1\right)}{16 \chi_0}\,, \\
  &\operatorname{l}_{2} =\frac{5}{128} \frac{-5 \sqrt{2} a_{-3} \left(73 \chi_0+100 \chi _1\right)+625 a_{-3}^2+5 \chi_0 \left(-15 \chi_0+120 \chi _1+224 \chi _2\right)}{128 \chi
   _0^2}\;, \\
	 & \cdots & \nn
\end{align}

\subsection{The FO-Theorem}\label{Sect:IV1}

 {\it This theorem relates the non-analyticity  of a function defined in a finite domain, to the large order behaviour of the coefficients of its Taylor expansion  at values where it is  analytic.}

In order to apply this theorem in our case  we first  project the domain $0\le X \le \infty$ to a finite one using the mapping:
\begin{equation}\lbl{eq:confXphi}
X \mapsto \varphi = \frac{1-X^2}{1+X^2}  \Longleftrightarrow X \mapsto \frac{\sqrt{1-\varphi}}{\sqrt{1+\varphi}} \Longleftrightarrow \begin{cases} X \rightarrow 0 \Longleftrightarrow \varphi \rightarrow 1 \\ X \rightarrow 1 \Longleftrightarrow \varphi \rightarrow 0 \\ X \rightarrow \infty \Longleftrightarrow \varphi \rightarrow -1\end{cases}
  \end{equation}
that projects $X$ to the domain {$\vert \varphi \vert \leqslant 1$}.
The FO-theorem is then encoded in the identity: 
\be\lbl{eq:FOid}
G_{\rm\tiny FO}\left(X = \frac{\sqrt{1-\varphi}}{\sqrt{1+\varphi}}\right) =  \sum_{n=0}^\infty\  \underbrace{(g_n -g_n^{\rm AS})}_{\cA_n}\ \varphi^n \ +\ \underbrace{\sum_{n=1}^\infty\  g_n^{\rm AS}\ \varphi^n}_{G_{\rm\tiny FO}^{\rm sing}(\varphi)}\,,
\ee
where the $g_n$ denote the coefficients of the Taylor expansion of $G_{\rm\tiny FO}(X)$ at $\varphi\ra 0$ and the $g_n^{\rm AS}$ the  coefficients of the same Taylor series as $n\ra\infty$. The FO-theorem relates the  $g_n^{\rm AS}$ coefficients to the  non-analyticity of the $G_{\rm\tiny FO}(X)$ function  at short distances ($\varphi\ra 1$) and at long-distances ($\varphi\ra -1$). The  second term in the r.h.s. of Eq.~\rf{eq:FOid} denotes the singular function $G_{\rm\tiny FO}^{\rm sing}(\varphi)$  that emerges from the sums of the asymptotic power series at $\varphi\ra 1$ and at $\varphi\ra -1$. 

More precisely, the Taylor expansion of the $G_{\rm FO}$ function at $X\rightarrow 1$ becomes now a Taylor  expansion at  $\varphi\ra 0$:  
\begin{equation}
  G_{\rm FO}\left(X=\frac{\sqrt{1-\varphi}}{\sqrt{1+\varphi}}\right) \underset{\varphi \rightarrow 0}{\sim} \sum_{n=0}^\infty g_n \varphi^n \,.
\end{equation}
Then:\begin{itemize}
	\item At short distances $ \varphi \ra 1$, and  from the expansion at $X\ra 0$ in Eq.~\rf{eq:FOseries}, one gets:  
\begin{multline}
\lbl{eq:expphiplus}
G_{\rm FO}\left({\frac{\sqrt{1-\varphi}}{\sqrt{1+\varphi}}}\right)  \underset{\varphi \rightarrow 1}{\sim}  \frac{\operatorname{s}_{2,0}}{2} (1-\varphi)+ \frac{2 \operatorname{s}_{3,0}-\ln 2 \operatorname{s}_{3,1}}{4 \sqrt{2}} (1-\varphi)^{\frac{3}{2}} \\
+ \frac{\operatorname{s}_{3,1}}{4 \sqrt{2}} (1-\varphi)^{\frac{3}{2}} \ln(1-\varphi) + \cdots \,.
\end{multline}
The second and third terms in this series  are at the origin of  the leading non-analytic contributions when $ \varphi\ra 1$. The FO-theorem relates them to the $n$-behaviour of  their contribution to the $g_n^{\rm AS}$ coefficients in  Eq.~\rf{eq:FOid}  as follows (see the Appendix for details):
{
\begin{equation}\lbl{eq:n1}
  (1-\varphi)^\frac{3}{2} \longmapsto \frac{2}{\sqrt{\pi}} \frac{1}{n^{\frac{5}{2}}} \left[1 + \frac{15}{8} \frac{1}{n} +\frac{385}{128} \frac{1}{n^2} + \cdots  \right]\;,
\end{equation}}
and
{ \setl
\bea
  \lefteqn{(1-\varphi)^\frac{3}{2} \ln(1-\varphi) \longmapsto} \nn \\
	& & \frac{2}{\sqrt{\pi}} \frac{1}{n^{\frac{5}{2}}} \left\{ \frac{8}{3}-\gamma_E-\ln 4 - \ln n  + \frac{15}{8}\left[\frac{56}{15}-\gamma_E-\ln 4 - \ln n \right] \frac{1}{n} + \cdots  \right\}\;. \lbl{eq:n2}
\eea}
The leading terms of these two asymptotic behaviours i.e., the term proportional to $\frac{1}{n^{5/2}}$ in Eq.~\rf{eq:n1} and the term proportional to $\frac{1}{n^{5/2}}\log n$ in Eq.~\rf{eq:n2},  generate then the following  singular functions: 
{ \be\lbl{eq:singshort}
\sum_{n=1}^\infty \frac{\varphi^n}{n^{5/2}}= \operatorname{Li}_{5/2}(\varphi)\quad \annd\quad
-\sum_{n=1}^\infty \frac{\log n}{n^{5/2}}\varphi^n=
\operatorname{Li}_{5/2}^{(1,0)}(\varphi)\,,
\ee }
where 
\be
  \operatorname{Li}^{(1,0)}_{a} \left(x\right) \doteq \frac{d}{d s}\operatorname{Li}_{s} \left(x\right)\bigg\vert_{s=a}\,.
\ee
These singular functions, modulated by their  corresponding coefficients, are then  to  be included in the function $G_{\rm FO}^{\rm sing}(\varphi)$  in Eq.~\rf{eq:FOid}.

\item At long distances $ \varphi \rightarrow -1$,  and  from the expansion at $X\ra \infty$ in Eq.~\rf{eq:FOseries}, one gets:
{ \begin{equation}\lbl{eq:expphiminus}
G_{\rm FO}\left(X=\frac{\sqrt{1-\varphi}}{\sqrt{1+\varphi}}\right)  \underset{\varphi \rightarrow -1}{\sim}  \frac{\operatorname{l}_{1}}{\sqrt{2}} \sqrt{1+\varphi} + \frac{\operatorname{l}_{2}}{2} (1+\varphi) + \frac{\operatorname{l}_{1}-2\operatorname{l}_{3}}{4 \sqrt{2}} (1+\varphi)^{\frac{3}{2}}+ \cdots \,.
\end{equation}}
The first term in the r.h.s. is at the origin of the leading non-analytic contribution when $ \varphi\ra -1$. The FO-theorem relates it to the $n$-behaviour of  its  contribution to the  $g_n^{\rm AS}$ coefficients in  Eq.~\rf{eq:FOid}  as follows (see the Appendix for details):
{ \begin{equation}
  \sqrt{1+\varphi} \longmapsto -\frac{(-1)^n}{2\sqrt{\pi}} \frac{1}{n^{\frac{3}{2}}} \left[1 + \frac{3}{8} \frac{1}{n} +\frac{25}{128} \frac{1}{n^2} + \cdots  \right]\,.
\end{equation}}
The leading term proportional to $\frac{1}{n^{3/2}}$ generates then the singular function
{ \begin{equation}\lbl{eq:singlong}
\sum_{n=1}^\infty \frac{(-1)^n\varphi^n}{n^{3/2}}= \operatorname{Li}_{3/2}(-\varphi)
\end{equation}}
that modulated by its corresponding coefficient, contributes to $G_{\rm FO}^{\rm sing}(\varphi)$  in Eq.~\rf{eq:FOid}.
\end{itemize}

 \subsection{ The FO-Approximants in Practice}\label{Sect:IV2}

A priori, the problem to implement in QCD the procedure discussed above  is that,  except for  the coefficients $a_{-3}$ and ${\chi}_0$, the other coefficients  of the asymptotic expansions     are not known  from first principles and, therefore,  practically all the coefficients $\operatorname{s}_{n,l}$ and $\operatorname{l}_{n}$ in Eqs.~\rf{eq:FOseries}  are unknown.  The FO-identity in Eq.~\rf{eq:FOid} and the  explicit  examples previously discussed   
show, however, the way to construct successive approximants to $G_{\rm FO}(X)$. The particular approximants that emerge from the leading non-analytic contributions discussed in the previous subsection are defined by 
  successive power series of  $N$ terms, plus a linear combination of the three types of singular functions in Eqs.~\rf{eq:singshort} and \rf{eq:singlong} i.e.,
\begin{multline}\lbl{eq:FOaps}
  G_{\rm FO}(X) \approx G_{\rm FO}^{(N;\frac{3}{2};\frac{5}{2},\frac{5^\prime}{2})}(X) = \sum_{n=0}^N \mathcal{A}_n \left(\frac{1-X^2}{1+X^2}\right)^n + \frac{\mathcal{B}_{-1,\frac{3}{2}}}{\eta(\frac{3}{2})} \operatorname{Li}_{\frac{3}{2}} \left(-\frac{1-X^2}{1+X^2}\right)\\
  + \frac{\mathcal{B}_{1,\frac{5}{2}}}{\zeta(\frac{5}{2})} \operatorname{Li}_{\frac{5}{2}} \left(\frac{1-X^2}{1+X^2}\right)+\frac{\mathcal{B}^{\prime}_{1,\frac{5}{2}}}{\zeta^\prime(\frac{5}{2})} \operatorname{Li}^{(1,0)}_{\frac{5}{2}} \left(\frac{1-X^2}{1+X^2}\right)\;,
\end{multline}
with coefficients
\be\lbl{eq:freeps}  
\mathcal{A}_n=g_n-g_n^\mathrm{AS}\,,\quad \mathcal{B}_{-1,\frac{3}{2}}\,,\quad  \mathcal{B}_{1,\frac{5}{2}}\,,
\ee
that are unknown parameters (they will be the free parameters in the fits discussed later);   exceptionally  
 the coefficient 
\be
\mathcal{B}^\prime_{1,\frac{5}{2}}=\frac{3}{16\sqrt{2\pi}}\ \zeta^\prime\left(\frac{5}{2}\right)\ \operatorname{s}_{3,1}\,,
\ee
is known 
because $\operatorname{s}_{3,1}$ given in Eq.~\rf{eq:s31} is fixed by $\chi
_0$~\footnote{In fact, all the coefficients $\mathrm{s}_{2n,n}$ for $n\ge 1$ depend only on $a_{-3}$ and $\chi_0$.}.
The polylog functions in Eq.~\rf{eq:FOaps} have been normalized, for convenience, to their values at $X=0$ where  $\zeta(s)$ denotes the Riemann zeta-function and $\eta(s)$, $\zeta^{\prime}(s)$ the related functions:
\begin{align}
&\eta (s) = \operatorname{Li}_s(-1) = \sum_{n=1}^\infty \frac{(-1)^n}{n^s}\,,\quad \zeta (s) = \operatorname{Li}_s(1) = \sum_{n=1}^\infty \frac{1}{n^s}\,,\nn\\
& \zeta^\prime (s) = \operatorname{Li}_s^{(1,0)}(1) = -\sum_{n=1}^\infty \frac{\ln n }{n^s}\,.
\end{align}
The parameters in Eq.~\rf{eq:freeps} are further restricted by the two sum rules:
{\setl
\bea\lbl{eq:srls}
0 & = & \sum_{n=1}^N \mathcal{A}_n +\mathcal{B}_{-1,\frac{3}{2}}+\mathcal{B}_{1,\frac{5}{2}}+ \mathcal{B}^\prime_{1,\frac{5}{2}}\,, \lbl{eq:SR1}\\
0 & = & \sum_{n=1}^N (-1)^n \mathcal{A}_n +\mathcal{B}_{-1,\frac{3}{2}}\ \frac{\zeta(\frac{3}{2})}{\eta(\frac{3}{2})}+\mathcal{B}_{1,\frac{5}{2}}\ \frac{\eta(\frac{5}{2})}{\zeta(\frac{5}{2})}+ \mathcal{B}^\prime_{1,\frac{5}{2}}\ \frac{\eta^\prime(\frac{5}{2})}{\zeta^\prime(\frac{5}{2})}\,, \lbl{eq:SR2} \\
{\rm where}\quad\eta^\prime (s) &  =  &  \operatorname{Li}_s^{(1,0)}(-1)= -\sum_{n=1}^\infty \frac{\ln n }{n^s}(-1)^n\,,
\eea}
which guarantee that the asymptotic behaviours 
\be
  \lim_{X\rightarrow 0}G_{\rm FO}(X)  = 0\quad\annd\quad \lim_{X\rightarrow \infty}G_{\rm FO}(X)  = 0
\ee
are satisfied.

The reason why  the polynomials in Eq.~\rf{eq:FOaps} are of finite   degree $N$  is due to the fact that only the contribution of the leading non-analytic terms  has been taken into account in the construction of the approximants. One expects, therefore, that beyond a certain critical $N$ value depending also on the input number of LQCD $G(x_0)$ values, the approximants will cease to improve. It is possible, however, to correct this by adding successive extra contributions associated to the subleading non-analytic terms, but it requires the introduction of extra $\cA_n$ parameters as well as further singular functions modulated by extra unknown $\cB$-like parameters and, furthermore,  a more refined and/or extended  set of  $G(x_0)$ input  values. In this work we shall, therefore, only consider the leading set of approximants defined in Eq.~\rf{eq:FOaps} which, as we shall see, already produce significantly  accurate results.

The values of the unknown parameters in Eq.~\rf{eq:freeps}, restricted to satisfy the two sum rules above,  can then be obtained from a linear fit of the successive $G_{\rm FO}^{(N;\frac{3}{2};\frac{5}{2},\frac{5^\prime}{2})}(X)$  approximants in Eq. \rf{eq:FOaps} to the data input provided by LQCD evaluations of $G(x_0)$ in a given optimal $x_0$-region. It is always possible to have a solution for the unknown parameters provided that:
  with $\boldsymbol{\mathcal{A}}=\{\mathcal{A}_0,\ldots,\mathcal{A}_N\}$ and
$\boldsymbol{\mathcal{G}}=\{\mathcal{G}(x_0),\ldots,\mathcal{G}(x_N)\}$ where
\begin{multline}
\mathcal{G}(\varphi_i) = G_{\mathrm{FO}}(\varphi_i) - \frac{\mathcal{B}^{\prime}_{1,\frac{5}{2}}\zeta^\prime(\frac{5}{2})}{2\zeta^2(\frac{5}{2})} \Bigg[ [c_1(0)-c_1(1)] \operatorname{Li}_{\frac{3}{2}} \left(-\varphi_i\right) + [c_2(0)-c_2(1)] \operatorname{Li}_{\frac{5}{2}} \left(\varphi_i\right) \\
+ 2\frac{\zeta(\frac{5}{2})}{\zeta^\prime(\frac{5}{2})} \operatorname{Li}^{(1,0)}_{\frac{5}{2}} \left(\varphi_i\right) \Bigg]\,,
\end{multline}
  the matrix  $\mathbb{M}$  such that $\mathbb{M}\cdot \boldsymbol{\mathcal{A}}= \boldsymbol{\mathcal{G}}$ is invertible. This implies the condition
\begin{equation*}
\det \mathbb{M}\equiv \det \left[\varphi_i^j + c_1(j) \operatorname{Li}_{\frac{3}{2}} \left(-\varphi_i\right) + c_2(j) \operatorname{Li}_{\frac{5}{2}} \left(\varphi_i\right) \right]_{i,j} \neq 0\,,
\end{equation*}
with
\be c_1(j)=\frac{4 \sqrt{2} (-1)^j-2+4 \sqrt{2}}{\left(\sqrt{2}-6\right) \zeta \left(\frac{3}{2}\right)}\quad \annd \quad c_2(j)=\frac{4 \left(\left(\sqrt{2}-1\right) (-1)^j+\sqrt{2}\right)}{\left(\sqrt{2}-6\right) \zeta \left(\frac{5}{2}\right)}\,,
\ee
 to be satisfied.

We illustrate in the next section how to implement this procedure with a phenomenological model that simulates the hadronic spectral function. 

\section{ Illustration with a Phenomenological  Model}\label{Sect:V}

The spectral function of the model in question  is  inspired from lowest order $\chi$PT, $\rho$-vector meson dominance, and asymptotic freedom:
\be
\lbl{eq:SFM}
\frac{1}{\pi}\Imm\Pi^{\rm HVP}_{\rm model}(t)=\frac{\alpha}{\pi}\left(1-\frac{4 m_{\pi}^2}{t}\right)^{3/2} \left\{ 
\frac{1}{12} \vert F(t)\vert^2 +
\sum_{\rm quarks}e_{q}^2\ \  \Theta(t,t_{c},\Delta)\right\}\theta(t-4m_{\pi}^2)\,.
\ee 

\begin{figure}[!ht]
\begin{center}
\includegraphics[width=0.7\textwidth]{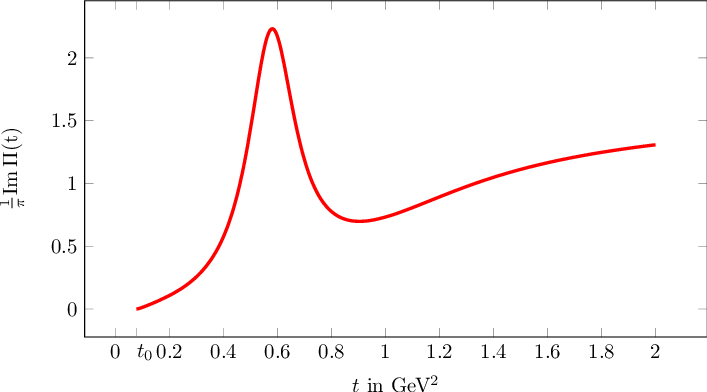} 
\caption{The model spectral function in Eq.~\rf{eq:SFM} for $t_c =1~\GeV^2$ and $\Delta=0.5~\GeV^2$ in $\frac{\alpha}{\pi}$-units.}
\lbl{fig:modelsp}
\end{center}
\end{figure}

\noi
It consists of a Breit-Wigner--like modulous squared  form factor~\footnote{This is a simplified version of  phenomenological spectral functions discussed in the literature, e.g. in refs.~\cite{PP,CHK21} and references therein.} 
\be\lbl{eq:ff2}
\vert F(t)\vert^2=\frac{M_{\rho}^4}{(M_{\rho}^2-t)^2 +M_{\rho}^2\  \Gamma(t)^2}\,,
\ee    
with an  energy dependent width
\be
\Gamma(t)=\frac{M_{\rho} t}{96\pi f_{\pi}^2}\left[\left(1-\frac{4 m_{\pi}^2}{t}\right)^{3/2}\theta(t-4m_{\pi}^2)+\frac{1}{2}\left(1-\frac{4 M_{k}^2}{t}\right)^{3/2}\theta(t-4M_{k}^2)
\right]\,;
\ee
plus a function
\be
\Theta(t,t_{c},\Delta)=\frac{\frac{2}{\pi}\arctan\left(\frac{t-t_{c}}{\Delta}\right)-\frac{2}{\pi}\arctan\left(\frac{t_{0}-t_{c}}{\Delta}\right)}{1-\frac{2}{\pi}\arctan\left(\frac{t_0-t_{c}}{\Delta}\right)}\,,
\ee
that has  two arbitrary parameters $t_c$ and $\Delta$  and smoothly matches the  low energy behaviour to  the asymptotic pQCD continuum. 
The shape of this spectral function, using the  physical central values for $m_{\pi}$, $M_k$,  $M_{\rho}$, $f_{\pi}=93.3~\MeV$, and the choice:  $t_c =1~\GeV^2$ and $\Delta=0.5~\GeV^2$, with   
$\sum_{\rm quarks}e_{q}^2=\frac{5}{3}$,  is shown in Fig~\rf{fig:modelsp}.

The shape of the TMR function of the model 
\be\lbl{eq:modelsp}
x_0^3\  G_{\rm model}(x_0)= { x_0^3}
\int_{\sqrt{t_0}}^\infty d\omega\   e^{-\omega{x}_0
}\ \omega^2 \frac{1}{\pi}\Imm\Pi^{\rm HVP}_{\rm model}(\omega^2)\,,
\ee
 is shown in Fig.~\rf{fig:Gmodel}.
The same  function plotted in terms of the variable $\varphi$ in Eq.~\rf{eq:confXphi} is shown in Fig.~\rf{fig:Gmodelphi}. Its contribution  to the muon anomaly, using only the center values of the parameters given above,  is
\be\lbl{eq:modelan}
\left(a_{\mu}^{\rm HVP}\right)_{\rm model}=6~992\times 10^{-11}\,.
\ee
\begin{figure}[!ht]
\begin{center}
\includegraphics[width=0.70\textwidth]{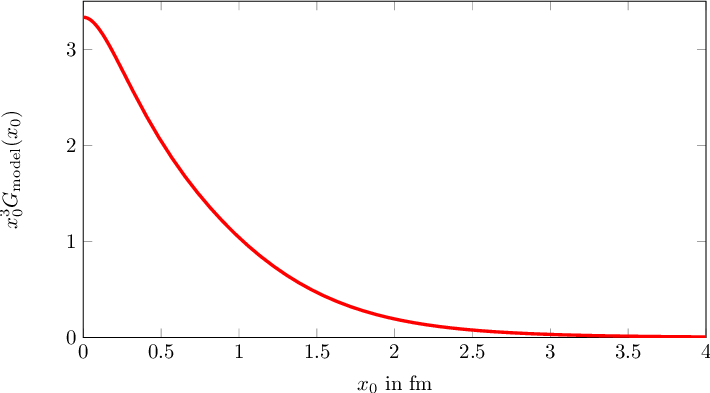} 
\caption{The model TMR function $x_0^3\ G_{\rm model}(x_0)$ in Eq.~\rf{eq:modelsp} in $\frac{\alpha}{\pi}$ units}
\lbl{fig:Gmodelphi}
\end{center}
\end{figure}
\begin{figure}[!ht]
\begin{center}
\includegraphics[width=0.70\textwidth]{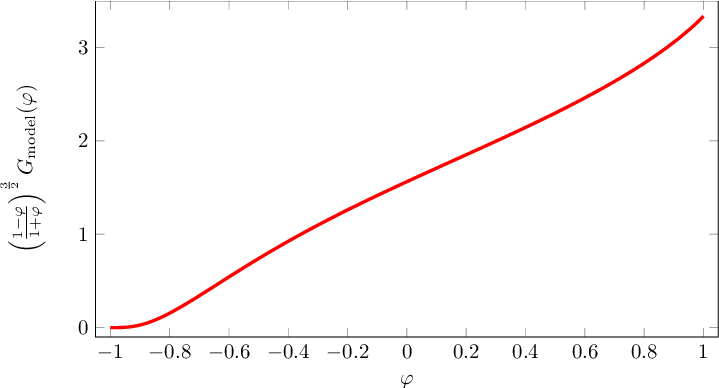} 
\caption{The model TMR function $x_0^3\ G_{\rm model}(x_0)$ as a function of the $\varphi$ variable in $\frac{\alpha}{\pi}$ units.}
\lbl{fig:Gmodel}
\end{center}
\end{figure}

We also show the shape of the integrand of $\left(a_{\mu}^{\rm HVP}\right)_{\rm model}$ as a function of the  $\varphi$ variable in Fig.\rf{fig:phiint}.  Notice that in this representation, the intermediate region $0.4{~\rm fm}\le x_0 \le 1.0~{\rm fm}$ favoured by the LQCD evaluations and shown in blue,  corresponds to the interval  $0.51\ge \varphi \ge -0.33$. One can see that, in spite of the exponential decrease of $G(x_0)$ at large $x_0$ (small $\varphi$), the contribution to $\left(a_{\mu}^{\rm HVP}\right)_{\rm model}$from the long-distance region $-1.0\ge \varphi\ge -0.33$ is highly weighted; a fact that demands a good reconstruction of $G(x_0)$ in the low-energy  region (large-$x_0$) in order  to have an accurate evaluation of $\left(a_{\mu}^{\rm HVP}\right)_{\rm model}$. This we expect to be a generic feature in QCD as well.
\begin{figure}[!ht]
\begin{center}
\includegraphics[width=0.8\textwidth]{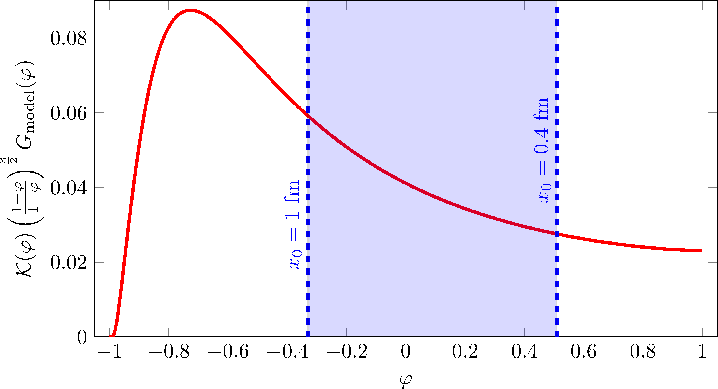} 
\caption{Shape of the integrand of $\left(a_{\mu}^{\rm HVP}\right)_{\rm model}$ in $\frac{\alpha}{\pi}\frac{m_{\mu}^2}{t_0}$ units as a function of $\varphi$}
\lbl{fig:phiint}
\end{center}
\end{figure}

The function
\be\lbl{eq:ratmodskel}
G_{\rm FO}^{\rm model}(x_0)\equiv \frac{G_{\rm model}(x_0)}{G^{*}(x_0)}-1\,,
\ee
with the two  parameters of the $G^{*}(x_0)$ function adjusted to the asymptotic behaviours of the model, i.e. $a_{-3}=\frac{10}{3}$ and $\chi_0 = 0.31$, 
is plotted in Figure~\rf{fig:FOmodel}  in black for a finite $x_0$ interval.  This shape is what  the successive FO-approximants in Eq.~\rf{eq:FOaps} are expected to reproduce, all the way from $x_0=0$ to $x_0=\infty$ (i.e. from $\varphi=1$ to $\varphi=-1$).  
The red dots in the figure are the points used in the fit described in Section~\rf{sec:fits}. More compact plots of the $G_{\rm FO}^{\rm model}$ function  in  terms of the $\varphi$-variable are shown in Fig.~\rf{fig:apphild} for $-1\le\varphi\le -0.5$ and in Fig.~\rf{fig:FOphiplotP} for $-0.5\le\varphi\le -1$.

\begin{figure}[!h]
\begin{center}
\includegraphics[width=0.75\textwidth]{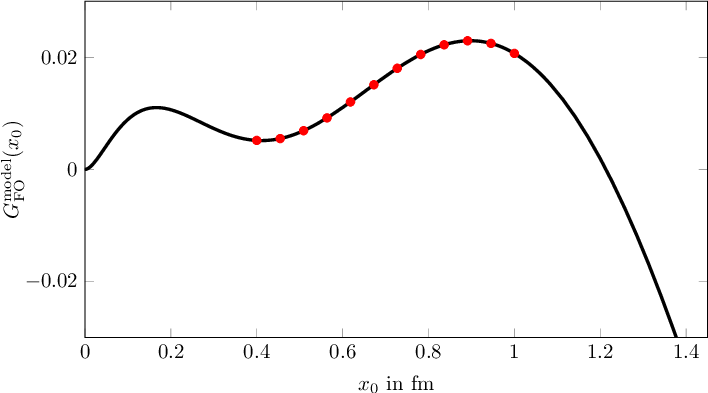} 
\caption{Shape  of the  $G_{\rm FO}^{\rm model}(x_0)$ function  in Eq.~\rf{eq:ratmodskel}     in the interval $0\le x_0 \le 1.4~{\rm fm}$.}
\lbl{fig:FOmodel}
\end{center}
\end{figure}
\begin{figure}[!h]
\begin{center}
\includegraphics[width=0.7\textwidth]{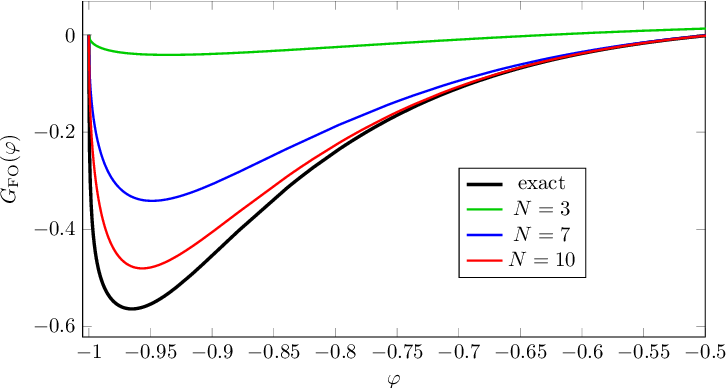} 
\caption{Shape  of the Approximants  $G_{\rm FO}^{(N;\frac{3}{2};\frac{5}{2},\frac{5^\prime}{2})}(\varphi)$  for $N=3$ (green), $N=7$ (blue) and $N=10$ (red). The model function  $G_{\rm FO}^{\rm model}(\varphi)$  which the approximants are expected to approach is in black.}
\lbl{fig:apphild}
\end{center}
\end{figure}
\begin{figure}[!h]
\begin{center}
\includegraphics[width=0.7\textwidth]{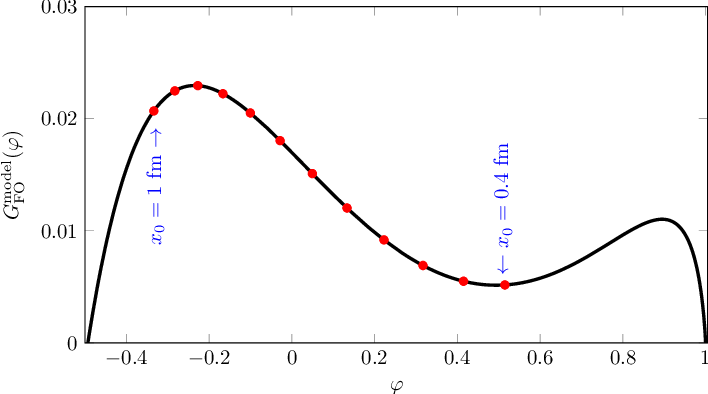} 
\caption{Shape  of the  $G_{\rm FO}^{\rm model}(\varphi)$ function    in the interval $-0.5\le\varphi\le 1$.}
\lbl{fig:FOphiplotP}
\end{center}
\end{figure}

 \subsection{Fits to the Model Data using FO-Approximants}\lbl{sec:fits} 

The input we use as an example are the values of the function  $G_{\rm FO}^{\rm model}(x_0)$ at twelve points, equally spaced  with no errors,  in the intermediate region~\footnote{This  is the $x_0$-region where at present  LQCD simulations  are  most precise~\cite{BMWmu,wittig,jansen,khadra}.}
\be
0.4~{\rm fm} \le x_0 \le 1.0~{\rm fm}\,.
\ee
The corresponding  data points are shown as red dots in Fig.~\rf{fig:FOmodel} and Fig.~\rf{fig:FOphiplotP}. We then make  linear fits  of the successive FO-approximants defined in Eqs.~\rf{eq:FOaps}, \rf{eq:SR1} and \rf{eq:SR2}  to the  $G_{\rm FO}^{\rm model}(x_0)$ function in Eq.~\rf{eq:ratmodskel}, and this way  obtain  the values of the   free parameters of each approximant that fix  the reconstruction of  the  $G_{\rm FO}^{\rm model}(x_0)$ function in the full $0\le x_0 \le\infty $ range. The corresponding reconstruction of $G(x_0)$ follows then from Eq.~\rf{eq:ratmodskel}.

The quality of the fits is shown in Fig.~\rf{fig:FOfits} for the approximants with $N=3$ in green, $  N=7$ in blue and $  N=10$ in red. The shape of $G_{\rm FO}^{\rm model}(x_0)$ is shown in black. One can see how  the reconstruction in the extended region $0.0~{\rm fm} \le x_0\le 1.5~{\rm fm}$ beyond  the one used for the fit,  improves as $N$ increases. The blue ($N$=7)  and red ($N$=10) curves are already quite closed to the $G_{\rm FO}^{\rm model}(x_0)$ black curve.

\begin{figure}[!h]
\begin{center}
\includegraphics[width=0.7\textwidth]{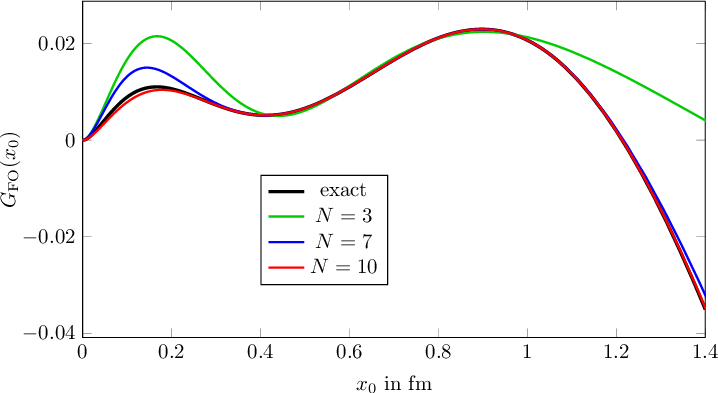} 
\caption{Shape  of the Approximants  $G_{\rm FO}^{(N;\frac{3}{2};\frac{5}{2},\frac{5^\prime}{2})}(x_0)$ in Eq.~\rf{eq:FOaps} for $N=3$ (green), $N=7$ (blue) and $N=10$ (red). The model function  $G_{\rm FO}^{\rm model}(x_0)$ which the approximants are expected to approach is in black.}
\lbl{fig:FOfits}
\end{center}
\end{figure}
In order to show the shapes of the approximants in the full $0\le X\le\infty$ it is better to use the  representation in terms of the equivalent 
 $\varphi$ variable, covering the full range $-1\le \varphi\le 1$. This is shown  in Fig.~\rf{fig:apphild} for $-1.0\le \varphi\le -0.5$ and, in a different scale, in Fig.~\rf{fig:apphisd} for $-0.5\le \varphi\le 1.0$.

The contribution of each approximant to the muon anomaly is then given by the integral
\begin{equation}\lbl{eq:anapps}
  a_\mu(N) = \int_0^\infty d X \, \mathcal{K}(X)\, X^3\, G^*(X)\left[1+G_{\rm FO}^{(N;\frac{3}{2};\frac{5}{2},\frac{5^\prime}{2})}(X)\right]\,,
\end{equation}
where $\mathcal{K}(X)$ is the kernel defined in Eq.~\rf{eq:kernel}
and $G_{\rm FO}^{(N;\frac{3}{2};\frac{5}{2},\frac{5^\prime}{2})}(X)$ the approximant defined in Eq.~\rf{eq:FOaps} with the values of the free parameters fixed by the fit. The results for each $N$-approximant  compared to the exact result in Eq.~\rf{eq:modelan} are given in Table~\rf{table:results}. The errors in \%  are the values of 
\begin{equation}
\lbl{eq:error}
  \mathtt{Err}(N) =   \left| \frac{a_\mu(N)-a_\mu^{\rm model}} {a_\mu(N)+a_\mu^{\rm model}} \right | 2 \times 10^2\,.
\end{equation}

\begin{figure}[!h]
\begin{center}
\includegraphics[width=0.8\textwidth]{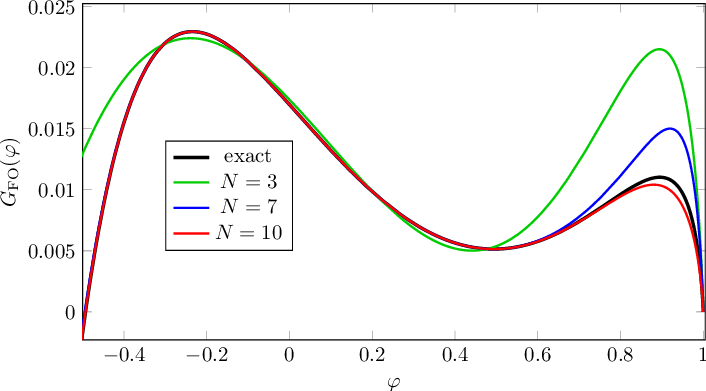} 
\caption{Shape  of the Approximants  $G_{\rm FO}^{(N;\frac{3}{2};\frac{5}{2},\frac{5^\prime}{2})}(\varphi)$ for $N=3$ (green), $N=7$ (blue) and $N=10$ (red) and their matching to the model function  $G_{\rm FO}^{\rm model}(\varphi)$ (black). The model function  $G_{\rm FO}^{\rm model}(\varphi)$  which the approximants are expected to approach is in black. The red and black curves in this region are already practically identical. Notice the vertical scale in the figure.}
\lbl{fig:apphisd}
\end{center}
\end{figure}

 \subsection{Errors of the FO-Approximants}\lbl{sec:errors}

\begin{table}[!h]
  \begin{center}
    \begin{tabular}{c||c|c}
   $N$ & $  a_\mu(N)$ in $10^{-11}$ units & $\mathtt{Err}(N)$ in $\%$\\
       \hline \hline
       1 & 7814 & 11 \\
   2 & 7696 & 9.6 \\
   3 & 7597 & 8.3\\
   4 & 7446 & 6.3 \\
   5 & 7335 & 4.8 \\
   6 & 7233 & 3.4 \\
   7 & 7162 & 2.4 \\
   8 & 7104 & 1.6 \\
   9 & 7066 & 1.0 \\
   10 & 7043 & 0.7 \\
   11 & 6990 & 0.04 \\
     \hline \hline
  \end{tabular}
\caption{Predicted  values of the model anomaly from the approximants defined in Eq.~\rf{eq:anapps} and the errors in \% defined by Eq.~\rf{eq:error} in the third column.}
\lbl{table:results}
\end{center}
\end{table}
\noi
The results in Table~\rf{table:results} show that  the approximants reproduce the value of $a_{\mu}^{\rm model}$ with better and better accuracy as $N$ increases. The best result is obtained for $N=11$ when the number of free parameters equals the number of input points and the linear fit corresponds then to solving a linear system of $N$ equations with $N$ unknowns. These results are very encouraging, however, 
in a potential application of the FO-Approximants to LQCD one will have to  take into account the errors of the input data  as well as an evaluation of the expected  error associated to the  FO-Approximants. Inclusion of the errors of the LQCD  data is beyond the scope of this paper~\footnote{This is something to  be discussed with each LQCD collaboration.},  
but two obvious questions which require answers on our part are:

\begin{enumerate}
	\item 
		{\it Given a finite number of $G(x_0)$ input values from LQCD simulations, and given the $N$  results of the successive reconstructions of the full $G(x_0)$ function using  FO-approximants,   what is the optimal value of $a_{\mu}^{\rm HVP}$ and what error should be assigned to it?}

\item 
{\it Can one give a  systematic error to the method of FO-Approximants?}
\end{enumerate}

An answer to the first question follows from the observation  in Table~\rf{table:results} that
\be
a_{\mu}(N+1)<a_{\mu}(N) \quad \foor~{\rm all}\quad N=1~{\rm to}~11\,.
\ee

\begin{itemize}
	\item

If this decreasing pattern persists in the case of an application to LQCD,  the value $a_{\mu}(N^{*})$ from the FO-approximant with $N=N^*$ i.e. the total number of input values,  is clearly the optimal choice. In this case it seems natural to assign as the error attributed to each $a_{\mu}(N)$ approximant the difference {  $\lvert a_{\mu}(N)-a_{\mu}(N^*)\lvert$. Provided that $N$ is sufficiently large, the optimal value  is then :
\be
a_{\mu}^{\rm optimal}=a_{\mu}(N^*)\pm \lvert a_{\mu}(N^{*}-1)-a_{\mu}(N^*)\lvert\,.
\ee}
\item
If the pattern of the $a_{\mu}(N)$ estimates, as $N$ increases, has a minimum or a maximum at a given $N^*$-value, then the optimal choice is the same as before  with $N^*$ at the value of the extrema.
\item
If the pattern  of the $a_{\mu}(N)$ approximants  oscillates as $N$ increases then  the most natural optimal  choice  is the one at the $N^*$ closest to the mean value of all the approximants. 
\end{itemize}

In order to get an estimate of the {\it systematic error} of the method of FO-approximants  when applied to a finite set  of input  values of $G(x_0)$, let us consider  the extreme case where $G(x_0)=G^{*}(x_0)$. The corresponding function $G_{\rm FO}(x_0)$  is then, by definition,  trivially zero.   However, because of the {\it systematic errors} of the FO-approximants, one expects  deviations from zero from the results of the approximants in the $x_0$ regions outside the one used as an input in the fit, and this is what one observes. Since in this case we know exactly the value of the muon anomaly (the one given by the chosen $G^{*}(x_0)$), we can define as a {\it systematic error} of each approximant  the one which follows from  applying the definition in Eq.~\rf{eq:error} to   this  case where $a_{\mu}(\rm model)=a_{\mu}^{*}$. We show  in Table~\rf{table:syst} this resulting systematic error for each  FO-approximant.  

\begin{table}[h!]
  \begin{center}
    \begin{tabular}{c|c}
   $N$ & $\mathtt{ErrorSyst}(N)$ in \%  \\
       \hline \hline
 1 & 1.00711 \\
 2 & 0.20572 \\
 3 & 0.16359 \\
 4 & 0.04585 \\
 5 & 0.05261 \\
 6 & 0.01495 \\
 7 & 0.02230 \\
 8 & 0.00484 \\
 9 & 0.01052 \\
 10 & 0.00070\\
 11 & 0.00524\\
\hline
\hline
  \end{tabular}\\ 
\caption{Systematic error in \% attributed to each FO-approximant defined in Eq.~\rf{eq:anapps}.}
\lbl{table:syst}
\end{center}
\end{table}

{ 
We have also analyzed the results of the FO-approximants in the following alternative situations:
\begin{itemize}
\item With only six input points, equally spaced, in the interval 0.4 fm to 1 fm one gets
\begin{equation}
a_\mu(5) = 7327 \times 10^{-11}\,
\end{equation}
which reproduces the model value at the level of $5\%$. The number of input points is however too small to give a significant systematic error in this case.

\item With an input of 10 points, equally spaced, but in the larger interval 0.3 fm to 1.2 fm as compared to the 0.4 fm to 1.0 fm interval used above, one gets
\begin{equation}
a_\mu(11) = 7003 \times 10^{-11}\,
\end{equation}
which reproduces the model value to $0.15\%$ with a systematic error of $0.2\%$:
\begin{equation}
a_\mu^\text{optimal} = (7003 \pm 15) \times 10^{-11}\,
\end{equation}
and indicates that using the same number of input points in a larger interval improves the result of the FO-approximants.

\end{itemize}
}
\section{ Conclusion and Outlook}\label{Sect:VI}

We have shown how the FO-theorem can be used to reconstruct the TMR function $G(x_0)$ in its full $0\le x_0 \le \infty$ domain, when one only uses as an input its values in a  restricted $x_0$-domain where  LQCD evaluations are   most precise. We have explicitly derived the functional  form of  the reconstruction  approximants that emerge from the properties of the FO-theorem.  These FO-approximants depend  linearly on  a set  of $N$ parameters that are related to the successive terms of the short-distance and long-distance expansions of the $G(x_0)$ function in QCD. The specific values of these QCD parameters are unknown, but they can be fixed from a fit of the FO-approximants to the LQCD evaluation of $G(x_0)$ in an optimal region.  In section \ref{Sect:V} we have illustrated  the procedure to follow in an eventual application to LQCD, with the simulation of a phenomenological model which captures the leading short and long distance behaviours of HVP in QCD. The application of FO-approximants in this case shows how the reconstruction of the model TMR function $G_{\rm model}(x_0)$ improves as the number $N$ of terms in the FO-approximant increases: using an input of twelve points, equally spaced with no errors, in the intermediate region $0.4~{\rm fm} \le x_0 \le 1.0~{\rm fm}$, we find that the best FO-approximant reproduces the value: $\left(a_{\mu}^{\rm HVP}\right)_{\rm model}=6~992\times 10^{-11}$ to an accuracy of $0.6$\%.  We find these results encouraging and worth considering for applications to the reconstruction of the $G(x_0)$ function in LQCD and the corresponding evaluations of $a_{\mu}^{\rm HVP}$ from first principles. 

Concerning the comparison of LQCD results with the data driven determinations in Eqs.~\rf{eq:disperevs} we suggest considering the case where the so called {\it skeleton function} introduced in Sections
\ref{Sect:III2} and \ref{Sect:IV} is chosen to be the one resulting from the data-driven determination of HVP. It is well known that the shape of this function is at present in disagreement with LQCD determinations in  intermediate $x_0$-windows~(see e.g. refs.~\cite{wittig}, \cite{khadra}, \cite{window}). The corresponding $G_{\rm FO}(x_0)$ function defined in Section \ref{Sect:IV} will, therefore,  be different from zero in these windows. The application of the FO-approximants in this case provides a way to evaluate how this difference propagates outside the region of $x_0$  used as an input. Comparing the optimal  $a_{\mu}^{\rm HVP}(N^{*})$ value obtained from the FO-approximants to the data driven results in Eqs.~\rf{eq:disperevs} would give an evaluation of the total discrepancy.

\acknowledgments{We thank J\'{e}r\^{o}me Charles for {his} participation at the early stages of this work. We are grateful to J\'{e}r\^{o}me Charles, Marc Knecht and { Harvey B. Meyer} for a careful reading of the manuscript and their comments.}

\newpage

\appendix

\vspace*{1cm}

\begin{center}
{\bf \Large Appendix}
\end{center}

\section{Mathematical details of the FO-theorem}

In full generality, the short-distance expansion in Eq.~\rf{eq:expphiplus} and the long-distance expansion in Eq.~\rf{eq:expphiminus} are given by the sums:
\begin{align}
G_{\rm FO}\left(X=\frac{\sqrt{1-\varphi}}{\sqrt{1+\varphi}}\right)  \underset{\varphi \rightarrow 1}{\sim} & \sum_{n\geqslant 2} \sum_{\ell \geqslant 0} \tilde{\operatorname{s}}_{\frac{n}{2},\ell} (1-\varphi)^{\frac{n+\ell}{2}} \ln^\ell (1-\varphi)\\
G_{\rm FO}\left(X=\frac{\sqrt{1-\varphi}}{\sqrt{1+\varphi}}\right)  \underset{\varphi \rightarrow -1}{\sim} & \sum_{n\geqslant 1} \tilde{\operatorname{l}}_{\frac{n}{2}} (1+\varphi)^{\frac{n}{2}}\;,
\end{align}
where the $\tilde{\operatorname{s}}_{\frac{n}{2},\ell}$ and  $\tilde{\operatorname{l}}_{\frac{n}{2}}$ coefficients are  linear combinations of the ${\operatorname{s}}_{\frac{n}{2},\ell}$ and ${\operatorname{l}}_{\frac{n}{2}}$   coefficients in Eqs.~\rf{eq:FOseries}.  The type of singular terms that appear in these expansions are: 

For $k$ and $\ell$  integer numbers, 
\begin{align}
 & (1-\varphi)^k \ln^\ell (1-\varphi) \lbl{eq:A3}\quad {\rm for}\quad \varphi\ra 1\,,  \\ & \nonumber \\
 & (1-\varphi)^\frac{2k+1}{2} \quad\text{and}\quad  (1-\varphi)^\frac{2k+1}{2} \ln^\ell (1-\varphi)\quad {\rm for} \quad\varphi\ra 1 \lbl{eq:A4}\,;
\end{align}
and  
\begin{equation}
(1+\varphi)^\frac{2k+1}{2} \quad {\rm for} \quad \varphi\ra -1 \,.\lbl{eq:A5}
\end{equation}
The FO-theorem gives 
the results for the large $n$ behaviour of the coefficients  $g_n^{\rm AS}$  in the $\varphi^n$ power series in Eq.~\rf{eq:FOid}, associated to these four types of singular terms. They can be found  in  the Appendix II  of ref.~\cite{GdeR22} and are given below.

\begin{itemize}
	\item 
 {\bf For $\ell=1$ in Eq.~\rf{eq:A3} } 
\begin{align}
  (1-\varphi)^k \ln\left(1-\varphi\right) & \longmapsto-\frac{(-1)^k\Gamma(k+1)}{n^{k+1}}\sum_{j=0}^\infty \begin{Bmatrix} k+j \\ m \end{Bmatrix}\frac{1}{n^j}\,,
\end{align}
where the symbol {\tiny $\begin{Bmatrix} \cdot \\ \cdot \end{Bmatrix}$} denotes Stirling numbers of the second kind.

\item {\bf For the first term  in Eq.~\rf{eq:A4}}

 The result can be directly obtained from the evaluation of the $\varphi^n$ coefficient of its Taylor series at  $\varphi\ra 0$, and then its behaviour as $n\ra\infty$:
\begin{equation}
\left[\varphi^n\right] \; \; (1-\varphi)^\frac{2k+1}{2} = \frac{\Gamma\left(-\frac{2k+1}{2}+n\right)}{\Gamma\left(-\frac{2k+1}{2}\right)\Gamma(n+1)} \underset{n\rightarrow \infty}{\sim} \frac{1}{n^{1+\frac{2k+1}{2}}} \sum_{j=0}^\infty \frac{\operatorname{B}_j^{[-\frac{2k+1}{2}]}(-\tfrac{2k+1}{2})}{\Gamma\left(-\frac{2k+1}{2}-j\right)\Gamma(1+j)}\frac{1}{n^j}\;,
\end{equation}
where  $\operatorname{B}^{[a]}_n(x)$ are the so-called \emph{generalized Bernoulli polynomials}~\cite{NIST:DLMF} or \emph{N\o rlund polynomials} (as encoded in \textsf{Mathematica} $\operatorname{B}^{[a]}_n(x) = \mathtt{Norlund}[n,a,x]$). Their first few terms are 
\begin{equation}
  \operatorname{B}^{[\lambda]}_0(\lambda)=1\;,\;  \operatorname{B}^{[\lambda]}_1(\lambda)= \frac{\lambda }{2}\;,\;  \operatorname{B}^{[\lambda]}_2(\lambda)= \frac{1}{12} \lambda  (3 \lambda -1)\;,\ldots\;.
\end{equation}

In particular, for $k=1$, this is the way that the result in Eq.~\rf{eq:n1} follows
\begin{equation}
  (1-\varphi)^\frac{3}{2} \longmapsto \frac{2}{\sqrt{\pi}} \frac{1}{n^{\frac{5}{2}}} \left[1 + \frac{15}{8} \frac{1}{n} +\frac{385}{128} \frac{1}{n^2} + \cdots  \right]\;.
\end{equation}

\item
{\bf For the second term in Eq.~\rf{eq:A4}}

We use the property that
\begin{equation}
(1-\varphi)^\frac{2k+1}{2} \ln^\ell(1-\varphi) = \frac{\partial^\ell}{\partial \varepsilon^\ell} \left[(1-\varphi)^{\frac{2k+1}{2}+\varepsilon}\right]_{\varepsilon=0}\;,
\end{equation}
where from the coefficient of the $\varphi^n$ term of its  Taylor series at $\varphi\ra 0$  can be easily calculated
\begin{align}
\left[\varphi^n\right] (1-\varphi)^\frac{2k+1}{2} \ln^\ell(1-\varphi) &= \frac{\partial^\ell}{\partial \varepsilon^\ell} \left[ \left[\varphi^n\right] (1-\varphi)^{\frac{2k+1}{2}+\varepsilon}\right]_{\varepsilon=0}\\
&=\frac{\partial^\ell}{\partial \varepsilon^\ell} \left[\frac{\Gamma\left(-\frac{2k+1}{2}-\varepsilon+n\right)}{\Gamma\left(-\frac{2k+1}{2}-\varepsilon\right)\Gamma(n+1)}\right]_{\varepsilon=0}\;.
\end{align}
Only the case $\ell=1$ is needed in our case with the result 
\begin{align}
&\left[\varphi^n\right] (1-\varphi)^\frac{2k+1}{2} \ln(1-\varphi) \nonumber\\
&= \frac{\Gamma\left(-\frac{2k+1}{2}+n\right)}{\Gamma\left(-\frac{2k+1}{2}\right)\Gamma(n+1)}\left[\psi\left(-\frac{2k+1}{2}\right)-\psi\left(n-\frac{2k+1}{2}\right)\right]\\
& \underset{n\rightarrow \infty}{\sim} \frac{-\ln n +\psi\left(-\frac{2k+1}{2}\right)}{n^{1+\frac{2k+1}{2}}} \sum_{j=0}^\infty \frac{\operatorname{B}_j^{[-\frac{2k+1}{2}]}(-\tfrac{2k+1}{2})}{\Gamma\left(-\frac{2k+1}{2}-j\right)\Gamma(1+j)}\frac{1}{n^j} - \frac{1}{n^{1+\frac{2k+1}{2}}} \sum_{j=0}^\infty \frac{\mathtt{b}_j\left(-\frac{2k+1}{2}\right)}{n^j} \;,
\end{align}
where the $\mathtt{b}_j\left(\lambda\right)$ are the polynomials
\begin{equation}
  \mathtt{b}_j\left(\lambda\right) = \frac{\delta_{j,0}}{\Gamma(\lambda)} + \sum_{m=1}^j \frac{\operatorname{B}_{j-m}^{[\lambda]}(\lambda)}{\Gamma\left(\lambda-j+m\right)\Gamma(1+j-m)} \frac{(-1)^m \operatorname{B}_{m}^{[1]}(\lambda)}{m}\,,
\end{equation}
and where we have also used the fact that
\begin{equation}
  \psi\left(n-\lambda\right) \underset{n\rightarrow \infty}{\sim} \ln n + \sum_{p=1}^\infty \frac{(-1)^p \operatorname{B}_{p}^{[1]}(\lambda)}{p}\frac{1}{n^p}\;.
\end{equation}
The result in Eq.~\rf{eq:n2} is the one which corresponds to the particular case where $\ell=1$ and $k=1$

{ \setl
\bea
  \lefteqn{(1-\varphi)^\frac{3}{2} \ln(1-\varphi) \longmapsto} \nn \\
	& & \frac{2}{\sqrt{\pi}} \frac{1}{n^{\frac{5}{2}}} \left\{ \frac{8}{3}-\gamma_E-\ln 4 - \ln n  + \frac{15}{8}\left[\frac{56}{15}-\gamma_E-\ln 4 - \ln n \right] \frac{1}{n} + \cdots  \right\}\;.
\eea}

 A general expression for any $\ell$ can be  easily obtained  from the results above before using successive derivatives in $\varepsilon$.

\item{\bf For the terms in Eq.~\rf{eq:A5}}

 The result can be directly obtained from the evaluation of the $\varphi^n$ coefficient of its Taylor series at  $\varphi\ra 0$, and then its behaviour as $n\ra\infty$:

{\setl
\bea
\left[\varphi^n\right] \; \; (1+\varphi)^\frac{2k+1}{2}  & = &  (-1)^n\frac{\Gamma\left(-\frac{2k+1}{2}+n\right)}{\Gamma\left(-\frac{2k+1}{2}\right)\Gamma(n+1)} \\
& \underset{n\rightarrow \infty}{\sim} & \frac{(-1)^n}{n^{1+\frac{2k+1}{2}}} \sum_{j=0}^\infty \frac{\operatorname{B}_j^{[-\frac{2k+1}{2}]}(-\tfrac{2k+1}{2})}{\Gamma\left(-\frac{2k+1}{2}-j\right)\Gamma(1+j)}\frac{1}{n^j}\,.
\eea}

\end{itemize}

The final expression for the $g^\mathrm{AS}_n$ coefficients follows  from the sum of the results given in the three items discussed above i.e.,
\begin{equation}
  g_n \underset{n\rightarrow \infty}{\sim} g^\mathrm{AS}_n = \frac{\mathcal{B}_{-1,\frac{3}{2}}}{\eta(\frac{3}{2})} \frac{(-1)^n}{n^{\frac{3}{2}}} + \frac{\mathcal{B}_{1,\frac{5}{2}}}{\zeta(\frac{5}{2})} \frac{1}{n^{\frac{5}{2}}} -\frac{\mathcal{B}^\prime_{1,\frac{5}{2}}}{\zeta^\prime(\frac{5}{2})}\frac{\ln n }{n^{\frac{5}{2}}}, 
\end{equation}
where the coefficient $\mathcal{B}_{-1,\frac{3}{2}}$ is proportional to $\mathrm{l}_1$ and the coefficients  $\mathcal{B}_{1,\frac{5}{2}}$ and $\mathcal{B}^\prime_{\frac{5}{2}}$ are a linear combination of the $\mathrm{s}_{n,\ell}$ coefficients.


\end{document}